\documentclass[preprint,aps,12pt,notitlepage,nofootinbib,tightenlines]{revtex4}
\usepackage{amsmath}
\usepackage{enumerate}
\usepackage{bm}
\usepackage{times}
\usepackage{braket}
\usepackage{color}
\usepackage{epsfig}
\usepackage{slashed}
\usepackage{hyperref}
\usepackage{multirow}
\usepackage{array}
\usepackage{float}
\usepackage{graphicx}
\usepackage{booktabs}
\usepackage[normalem]{ulem} %下划线
\usepackage{adjustbox}
\newcommand{\beq}{\begin{eqnarray}}
\newcommand{\eeq}{\end{eqnarray}}
\newcommand{\be}{\begin{equation}\begin{aligned}}
\newcommand{\ee}{\end{aligned}\end{equation}}

\definecolor{Red}{rgb}{1.,0.,0.}

\definecolor{Blue}{rgb}{0.,0.,1.}

\definecolor{nicered}{rgb}{0.7,0.1,0.1}
\definecolor{nicegreen}{rgb}{0.1,0.5,0.1}
\def\lsim{ {\ \lower-1.2pt\vbox{\hbox{\rlap{$<$}\lower6pt\vbox{\hbox{$\sim$}}}}\ } }
\def\gsim{ {\ \lower-1.2pt\vbox{\hbox{\rlap{$>$}\lower6pt\vbox{\hbox{$\sim$}}}}\ } }

\hypersetup{colorlinks,citecolor=nicegreen,linkcolor=nicered}
\begin{document}

\title{Searching for single production of a vector-like $Y$ quark \\ decaying into $bW$ at the FCC-eh}
\author{Liangliang Shang$^{1}$\footnote{E-mail: shangliangliang@htu.edu.cn; liangliang.shang@physics.uu.se}}
\author{Weifan Zhu$^{1}$\footnote{E-mail: zhuweifan@stu.htu.edu.cn}}
\author{Bingfang Yang$^{1}$\footnote{E-mail: yangbingfang@htu.edu.cn. Corresponding author.}}
\author{Stefano Moretti$^{2,3}$\footnote{E-mail: s.moretti@soton.ac.uk; stefano.moretti@physics.uu.se}}

\affiliation{1. School of Physics, Henan Normal University, Xinxiang 453007, PR China\\ 
2. School of Physics \& Astronomy, University of Southampton, Highfield, Southampton SO17 1BJ, UK \\
3. Department of Physics \& Astronomy, Uppsala University, Box 516, 751 20 Uppsala, Sweden}
%%\date{\today}

\begin{abstract}
    We investigate the exclusion and discovery potential for single production of a vector-like $Y$ quark with electric charge $Q=-4/3$, followed by the decay $Y\to bW$, at the FCC-eh. The $Y$ quark is allowed to couple to both first- and third-generation down-type quarks. The analysis is performed for an electron-beam polarization of $P_e=-80\%$ at $\sqrt{s}=3.46$, $5.29$, and $6.9~\mathrm{TeV}$. Both leptonic and hadronic $W$-boson decay channels are considered. In the hadronic channel, the boosted $W$-boson is reconstructed as a $W$-jet, and kinematic observables are used to suppress the Standard Model (SM) backgrounds. By performing a detailed detector simulations and event analysis, we present the $2\sigma$ exclusion limits and $5\sigma$ discovery reaches in the $g^*$--$m_Y$ plane, where $g^*$ is 
    $Y$ coupling strength to the SM quarks.
    We find that the hadronic channel can provide stronger exclusion and discovery sensitivities, which are improved with increasing $\sqrt{s}$ at the FCC-eh.
\end{abstract}
\maketitle  %写上这个，论文题目，机构作者才会显示
%%====================================================================
\newpage
\section{Introduction}
The Standard Model (SM) has achieved remarkable success in describing the electroweak and strong interactions of elementary particles. In particular, the discovery of the Higgs boson in 2012 confirmed the mechanism of electroweak symmetry breaking~\cite{ATLAS:2012yve,CMS:2012qbp}. Nevertheless, the SM leaves several open questions, such as the quadratic sensitivity of the Higgs mass parameter, the origin of fermion masses and flavor hierarchy, and the generation of neutrino masses~\cite{Arkani-Hamed:2026wwy}. These issues motivate physics beyond the Standard Model (BSM). Among these, the hierarchy problem is closely related to the large radiative corrections to the Higgs mass parameter induced by the top quark through its Yukawa coupling. In many BSM scenarios, new fermionic states, commonly referred to as top partners, can soften these corrections and reduce the degree of fine-tuning~\cite{Burdman:2014zta}.

A sequential fourth generation of chiral quarks would obtain its mass through the Higgs mechanism, as in the case of the known SM fermions~\cite{He:2001tp,Kribs:2007nz}. Their large Yukawa couplings would induce sizable non-decoupling effects in loop-induced Higgs production, especially in the gluon-fusion process $gg\to H$. As a result, the fourth chiral generation would enhance the Higgs production rate by roughly a factor of nine relative to the SM prediction. This result is in strong tension with Large Hadron Collider (LHC)  measurements~\cite{Lenz2013} and is also severely constrained by the electroweak oblique parameters~\cite{Erler:2004nh}.

In contrast, vector-like quarks (VLQs) contain left- and right-handed components with identical gauge quantum numbers. Their masses can therefore arise from gauge-invariant Dirac mass terms independently of electroweak symmetry breaking. As a result, VLQs can avoid the severe non-decoupling effects associated with a sequential chiral generation. Recently, a TeV-scale flavor framework provides an additional theoretical motivation for vector-like fermions~\cite{Arkani-Hamed:2026wwy}. In this class of models, direct SM Yukawa couplings for light fermions are forbidden by a flavor symmetry and are generated effectively through light-heavy Yukawa interactions and soft flavor-symmetry-breaking mass terms after the heavy vector-like states are integrated out. The fermion mass hierarchy is encoded in chain-like structures, and the same chain structure can also be extended to the neutrino sector, where singlet states at the endpoints of the chains allow Majorana mass terms and generate parametrically suppressed light-neutrino masses.

According to their representations under $SU(2)_L$, VLQs can appear as singlets, doublets, or triplets. Their common members include the $T$ quark with electric charge $+2/3$, the $B$ quark with charge $-1/3$. Exotic states $X$ and $Y$ with charges $+5/3$ and $-4/3$, respectively, also appear in higher representations. Specifically, the $Y$ quark can appear only in doublet or triplet representations~\cite{Shang:2024wwy}. In realistic composite Higgs models, tree-level custodial protection of the $T$ parameter and the ratio $R_b$ usually requires several VLQ multiplets. If the scalar sector contains only $SU(2)_L$ doublets, as in the SM, VLQs with renormalizable couplings to SM quarks can appear only in seven gauge-covariant multiplets, including the $(B,Y)$ doublet and the $(T,B,Y)$ triplet~\cite{Aguilar-Saavedra:2013qpa}. Phenomenologically, an exotic $Y$ quark can decay into a $W$-boson and a down-type quark ($d$, $s$, or $b$), leading to distinctive collider signatures~\cite{Aguilar-Saavedra:2013wba}. Several studies of $Y$-quark production and decay have been performed~\cite{Arhrib:2024tzm,Shang:2024wwy,Han:2025itd,CETINKAYA2021115580}.

In many studies, the $Y$ quark is assumed to couple predominantly to third-generation quarks, because the couplings to first-generation quarks are strongly constrained by atomic parity violation (APV) measurements and may induce tree-level flavor-changing neutral currents (FCNCs)~\cite{BRANCO1986738,DELAGUILA1983107}. Nevertheless, allowing VLQs to couple to the first-generation quarks can enhance the production cross section and may also provide a mechanism for generating the observed fermion mass hierarchy. If a flavor symmetry $G$ forbids direct Yukawa couplings between the SM light quarks ($u$, $d$, $s$, $c$) and the Higgs field, their masses can be generated dominantly through mixing with heavy VLQs. By adjusting the mixing strengths between VLQs and different SM generations, a large hierarchy among the light-quark masses can be obtained without requiring extremely small Yukawa coefficients~\cite{Botella:2016ibj}.

At high-energy colliders, $Y$ quarks can be produced either in pairs or singly. Pair production proceeds through QCD interactions and remains largely model-independent, but it suffers significant phase-space suppression for large VLQ masses. ATLAS has set the most stringent pair-production limit for a $Y$ quark decaying exclusively into $bW$, excluding masses below $m_Y=1.7~\mathrm{TeV}$ in final states with leptons and jets using $140~\mathrm{fb}^{-1}$ of data at $\sqrt{s}=13~\mathrm{TeV}$~\cite{2024138743}. And CMS has excluded $Y$ quark masses below $1.295~\mathrm{TeV}$ in the $Y\to bW$ channel using $35.8~\mathrm{fb}^{-1}$ of data at $\sqrt{s}=13~\mathrm{TeV}$ \cite{CMS_Y_pairduction}.

Single production depends on the mixing parameters between the VLQs and SM quarks. It proceeds through electroweak interactions and is therefore model-dependent. Compared with pair production, single production is less suppressed by phase-space at large VLQ masses, although its rate is proportional to the square of the relevant mixing parameter. With increasing lower limits placed on VLQ masses~\cite{Aguilar-Saavedra:2013qpa}, single production becomes increasingly important. For single production of a $Y$ quark decaying into $bW$, ATLAS has examined several final states in the $(B,Y)$ doublet scenario, including $Y\to b \ell^- \nu$~\cite{ATLAS:2018dyh} and $Y\to b j j$~\cite{ATLAS:2024kgp}. In the leptonic channel, $\kappa\geq0.17$ are excluded for $m_Y=0.8$--$1.2~\mathrm{TeV}$ in the $(B,Y)$ doublet model~\cite{ATLAS:2018dyh}. In the hadronic channel, upper limits on $\kappa$ in the range $0.3$--$0.7$ are obtained for $m_Y=2.0$--$2.4~\mathrm{TeV}$, while values of $\kappa\geq0.24$ are excluded around $m_Y=1.5~\mathrm{TeV}$~\cite{ATLAS:2024kgp}. CMS has set the strongest exclusion limit for single production of VLQs decaying into $bW$, which excludes $\kappa\geq0.2$ for $m_Y=0.8$--$2.4~\mathrm{TeV}$~\cite{CMS:2026vwc}.

Hadron-hadron colliders provide high $\sqrt{s}$ and large luminosities, but they also suffer from substantial QCD backgrounds. Lepton-lepton colliders offer cleaner experimental environments but are limited in the direct production reach for heavy states. In contrast, Lepton-hadron colliders can provide a relatively clean environment and can reach high $\sqrt{s}$. A prominent example is the electron-hadron Future Circular Collider (FCC-eh), with electron-beam energies of $E_e=60~\mathrm{GeV}$ (design scenario)~\cite{Abada2019FCC}, $140~\mathrm{GeV}$ (possible scenario)~\cite{LHeC:2020van,LHeCStudyGroup:2012zhm,Acar:2016rde}, and $240~\mathrm{GeV}$ (optimistic scenario)~\cite{Curtin2018LLP}, together with a fixed proton-beam energy of $E_p=50~\mathrm{TeV}$. These configurations correspond to $\sqrt{s}=3.46$, $5.29$, and $6.9~\mathrm{TeV}$, respectively. The FCC-eh is designed to deliver an integrated luminosity of about $2000~\mathrm{fb}^{-1}$~\cite{Abada2019FCC,Bordry:2018gri}. In addition, electron-beam polarization provides an important handle for enhancing charged-current processes and probing electroweak interactions. At such lepton-hadron colliders, single production is expected to be the dominant production mode for heavy $Y$ quarks.

Motivated by these considerations, we study single production process $e^-p\to Y\nu_e$ followed by $Y\to bW$ at the FCC-eh. We allow the $Y$ quark to couple to both first- and third-generation down-type quarks and consider the leptonic channel, $Y\to bW (\to \ell^- \nu)$, and the hadronic channel, $Y\to bW (\to j j)$. In addition, the polarization of the electron-beam is also taken into account~\cite{10.3389/fphy.2022.886473}. 

This paper is organized as follows. In Sec.~\ref{section2}, we introduce the effective Lagrangian and describe the simulation setup. In Sec.~\ref{section3}, we discuss the event analysis for the two decay channels. In Sec.~\ref{section4}, we present the exclusion limits and discovery reaches. Finally, we present the conclusions in Sec.~\ref{section5}.

\section{Simplified model for a doublet $Y$ quark}
\label{section2}

\subsection{Effective Lagrangian for a doublet $Y$ quark}

Due to the stronger flavor constraints on the $Y$ coupling to the second generation SM quark \cite{Buchkremer:2013bha}, we consider a simplified scenario in which the vector-like $Y$ quark couples only to the first and third generations. The effective Lagrangian describing the interactions between the $Y$ quark and the SM quarks via the $W$-boson is given below (only include the terms relevant to the simulation and analysis):
\begin{equation}
\mathcal{L}_Y = g^* \left\{
    \sqrt{\frac{R_L}{1+R_L}} \frac{g}{\sqrt{2}} \left[ \bar{Y}_R W_\mu^- \gamma^\mu d_R \right]
    + \sqrt{\frac{1}{1+R_L}} \frac{g}{\sqrt{2}} \left[ \bar{Y}_R W_\mu^- \gamma^\mu b_R \right]
    \right\} + \mathrm{H.c.} 
    \label{Y}
\end{equation}
\noindent Here, $g$ is the $SU(2)_L$ gauge coupling, $g^*$ denotes the overall coupling strength of the $Y$ quark to SM quarks, and $R_L$ is the generation-mixing parameter. The squared factors $R_L/(1+R_L)$ and $1/(1+R_L)$ determine the relative coupling strengths to first- and third-generation quarks, respectively. In the limiting cases $R_L=0$ and $R_L\to\infty$, the $Y$ quark couples exclusively to third- and first-generation quarks, respectively. For singlet and triplet VLQs, the left-handed mixing typically dominates, whereas for doublet representations the right-handed mixing is dominant. Since the $Y$ quark considered in this work belongs to the $(B,Y)$ doublet, we retain only the right-handed interactions in Eq.~(\ref{Y})~\cite{Buchkremer:2013bha,Cao:2015doa,Cetinkaya:2020yjf}. The partial width for $Y\to bW$ is:
\begin{equation}
    \Gamma(Y \to bW) \simeq
    \frac{\alpha_e {g^*}^2 (m_Y^2-m_W^2)^2(2m_W^2+m_Y^2)}
    {16s_W^2m_W^2m_Y^3(1+R_L)} .
\end{equation}
Here, $\alpha_e=g^2/(4\pi)$, $s_W\equiv\sin\theta_W$ and the $b$ quark mass is neglected.

When the $Y$ quark mixes with SM quarks, constraints from the oblique parameters $S$, $T$, and $U$ should be considered. In the $(B,Y)$ doublet model, the corresponding contributions can be approximated as~\cite{Cao_2022}:
\begin{equation}
    S \simeq \frac{1}{2\pi} \left\{ -\frac{2}{3} {g^*}^2 \ln \frac{\mathcal{M}^2}{m_b^2} + \frac{11}{3} {g^*}^2 \right\}, \quad 
    U \simeq -\frac{{g^*}^2}{2\pi}, \quad 
    T \simeq \frac{3m_t^2}{8\pi \sin^2 \theta_W m_W^2} {g^*}^4 \frac{2\mathcal{M}^2}{3m_t^2}.
    \label{equationSTU}
\end{equation}
Here, $\mathcal{M}^2=m_Y^2/(1-{g^*}^2)$, $m_W=m_Z\cos\theta_W$, and $g^*=\sin\theta_R^b$ in our simplified model. For the numerical analysis, we require the oblique-parameter fit to satisfy $\chi^2<8.02$, corresponding to the $2\sigma$ region for three degrees of freedom. The fit values used in this work are $S=0.021\pm0.096$, $T=0.04\pm0.12$, and $U=0.008\pm0.092$~\cite{ParticleDataGroup:2024cfk}. For example, Eq.~(\ref{equationSTU}) gives the approximate upper bound $g^*<0.2$ for $m_Y=1500~\mathrm{GeV}$.

If the $Y$ quark couples to first-generation SM quarks, mass-independent constraints can be derived in the $g^*$--$R_L$ plane. The most relevant constraint arises from deviations in APV observables~\cite{Deandrea:1997wk,Arcadi:2019uif,Moretti:2025ckw}. The weak charge of a nucleus can be written as~\cite{Deandrea:1997wk}:
\begin{equation}
    Q_W = \frac{2\cos\theta_W}{g}
    \left[(2Z+N)(g_{ZL}^{u}+g_{ZR}^{u})+(Z+2N)(g_{ZL}^{d}+g_{ZR}^{d})\right],
    \label{Q_W}
\end{equation}
where $Z$ and $N$ denote the proton and neutron numbers, respectively. The quantities $g_{ZL}^{u,d}$ and $g_{ZR}^{u,d}$ denote the left- and right-handed couplings of the $Z$-boson to up- and down-type quarks. Since the doublet $Y$ quark couples to SM down-type quarks and the left-handed component of the VLQ doublet is neglected, one obtains a correction to the weak charge of the form:
\begin{equation}
    \delta Q^{Y}_{W} \simeq 2(Z+2N)T^B_3 |(V_R^b)^{41}|^2 ,
    \label{deltaQY}
\end{equation}
where $T_3^B$ denotes the weak isospin of the vector-like $B$ quark in the $(B,Y)$ doublet, and $|(V_R^b)^{41}|$ parameterizes the mixing between the VLQ sector and the first-generation down-type quark. The most precise APV measurements are obtained from Cesium $^{133}\mathrm{Cs}$ ($Z=55$, $N=78$) and Thallium $^{204}\mathrm{Tl}$ ($Z=81$, $N=123$)~\cite{PhysRevD.110.030001}. Using Eq.~\eqref{deltaQY}, the corresponding $2\sigma$ constraints on the doublet $Y$ representation can be expressed as:
\begin{equation}
    |\delta Q_W^Y| = (Z+2N)|(V_R^b)^{41}|^2
  \quad \Rightarrow \quad
  \begin{cases}
  0.0068 < |(V_R^b)^{41}| < 0.089, & ^{133}\mathrm{Cs}\ \mathrm{APV}, \\
  |(V_R^b)^{41}| < 0.15, & ^{204}\mathrm{Tl}\ \mathrm{APV}.
  \end{cases}
  \label{APV}
\end{equation}

The Cesium upper limit is more stringent than the Thallium one. In our simplified model, using the Cesium upper bound in Eq.~\eqref{APV} and $|(V_R^b)^{41}|=g^*\sqrt{\frac{R_L}{1+R_L}}$, one obtains $g^*\lesssim0.13$ for $R_L=1$, while the constraint becomes weaker for smaller $R_L$. Conversely, for $g^*=0.5$, the APV bound requires $R_L\lesssim0.03$, which strongly suppresses the first-generation coupling. Guided by the parameter ranges commonly adopted in VLQ phenomenological studies~\cite{SHANG2022115977,Zhang2024,Yang2024,Liu:2024hvp,Han2023,Han2022}, we adopt the scan ranges $g^*\leq0.5$ and $0\leq R_L\leq1$ since we are interested in the impact of first-generation couplings.

\subsection{Simulation and event generation}

\begin{figure}[!t]
    \centering
    \includegraphics[width=0.45\textwidth]{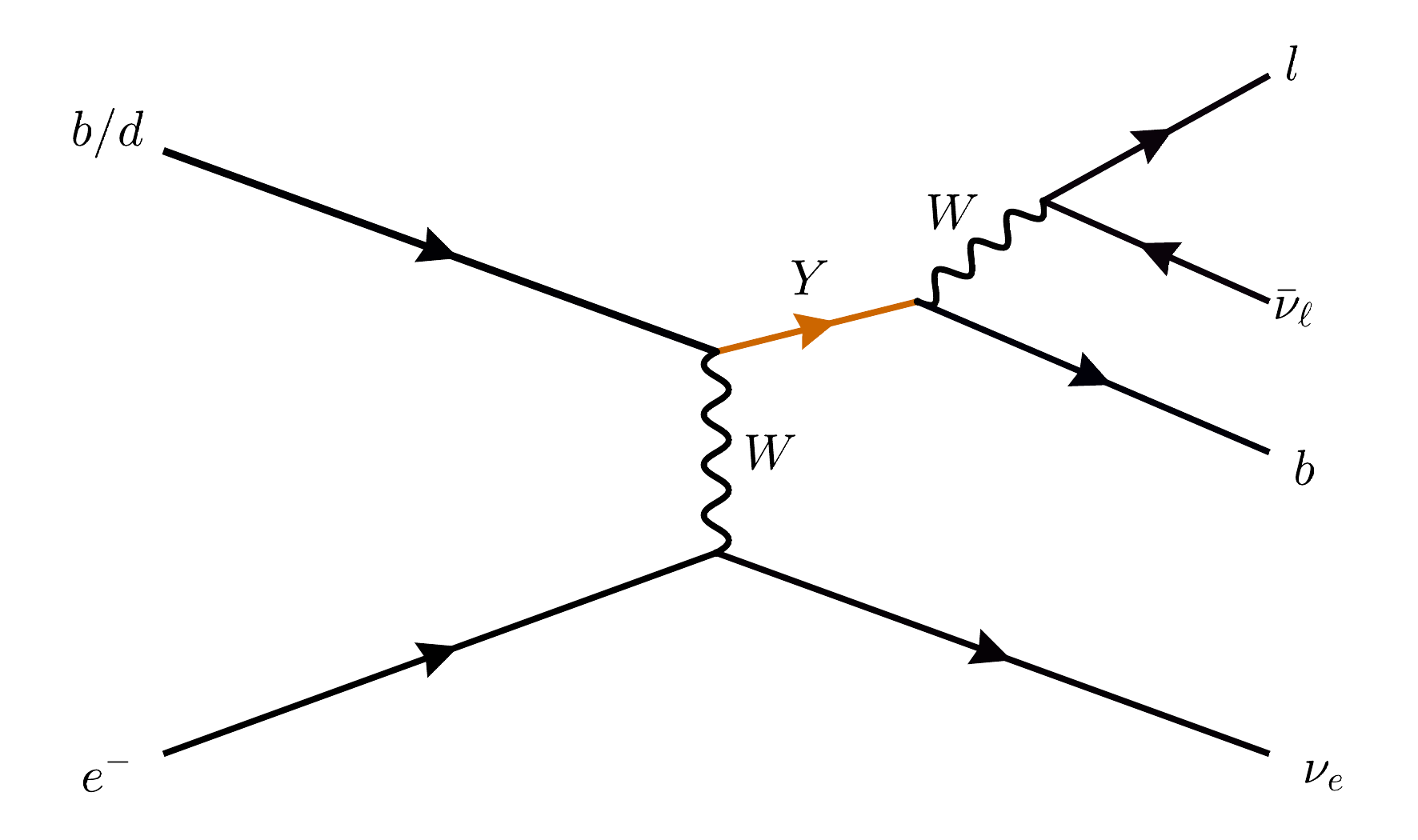}%
    \hspace{0.1\textwidth}%
    \hspace{-1cm}%
    \includegraphics[width=0.45\textwidth]{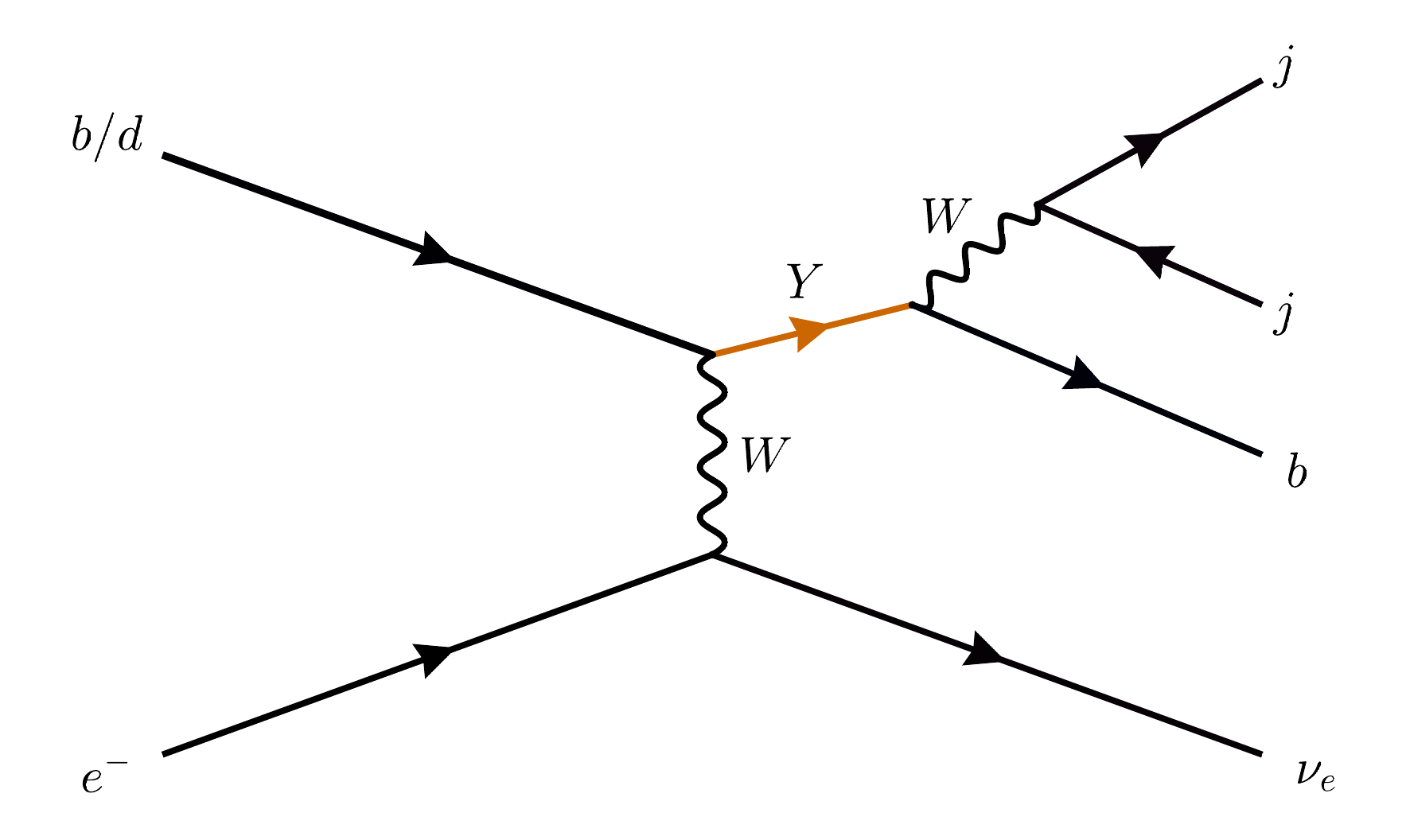} 
    \caption{Representative Feynman diagrams for single production of the $Y$ quark at the FCC-eh in Case~1 (left) and Case~2 (right).}
    \label{fig3}
\end{figure}

We analyze the leptonic and hadronic decay channels of the $W$-boson in single $Y$-quark production, which lead to two distinct signal signatures. The corresponding Feynman diagrams are shown in Fig.~\ref{fig3}, and the five-flavor scheme is adopted. The leptonic channel, referred to as Case~1 (see Fig.~\ref{fig3}, left),
\begin{align*}
    e^-p \to Y(\to bW)\nu_e \to bW(\to \ell^- \bar{\nu}_{\ell})\nu_e .
\end{align*}
and the hadronic channel, referred to as Case~2 (see Fig.~\ref{fig3}, right), 
\begin{align*}
    e^-p \to Y(\to bW)\nu_e \to bW(\to jj)\nu_e .
\end{align*}
Here, $\ell^- = e^-,\mu^-$. 

%%%%%%%%%%%%%%%%%%%%%%%%%%%%%%%%%%%%%%%%
%%%%%%%%%%%%%%%%%%%%%%%%%%%%%%%%%%%%%%%%%%%%

For Case~1, the main SM background processes are $e^- p \to \bar{t} (\to \bar{b} W)\nu_e  \to \bar{b} W (\to \ell^-  \bar{\nu}_{\ell}) \nu_e$, $e^- p \to W (\to \ell^- \bar{\nu}_{\ell}) j \nu_e$, and $e^- p \to e^- Z (\to \nu \bar{\nu}) j$.
For Case~2, the main SM background processes are
$e^- p \to W (\to j j) j \nu_e$,
$e^- p \to \bar{t} (\to \bar{b} W) \nu_e \to \bar{b} W (\to j j)\nu_e$, and $e^- p \to Z (\to j j) j \nu_e$.

The numerical values of the input SM parameters are taken as follows~\cite{ParticleDataGroup:2024cfk}:
\begin{align*}
m_b = 4.183~&\mathrm{GeV}, \quad
m_t = 172.57~\mathrm{GeV}, \quad
m_Z = 91.1880~\mathrm{GeV}, \quad
m_W = 80.3692~\mathrm{GeV}, \\
&\sin^2\theta_W = 0.22321, \quad
\alpha(m_Z) = 1/127.930, \quad
\alpha_s(m_Z) = 0.1180.
\end{align*}

We impose the following basic cuts at the parton level:
\begin{equation*}
\Delta R(x,y)>0.4 \,(x,y=\ell,j),\quad p_T(j/b)>20~\mathrm{GeV},\quad |\eta(j/b)|<5, \quad p_T(\ell)>10~\mathrm{GeV},\quad |\eta(\ell)|<5 .
\end{equation*}
Here, $\Delta R$ denotes the separation in the pseudorapidity-azimuth plane. 

Hard-scattering cross sections and signal/background events are generated with MadGraph5\_aMC@NLO~3.4.2~\cite{MadGraph} using the NNPDF23LO1 parton distribution functions (PDFs)~\cite{NNPDF}. 
The default dynamic factorization and renormalization scales of MadGraph5\_aMC@NLO are used in our simulatioin. The parton-level events are showered and hadronized with PYTHIA~8.3~\cite{Pythia}, and fast detector simulation is performed with Delphes~3.5~\cite{Delphes}. Small-radius jets are reconstructed using the anti-$k_t$ algorithm~\cite{Antikt}, implemented in FastJet~3.3.4~\cite{FastJet}, with radius parameter $R=0.4$. The $b$-tagging efficiency is taken to be $85\%$ for $|\eta|<2.5$ and $64\%$ for $2.5\leq|\eta|<4.0$, with a mild $p_T$-dependent degradation at high transverse momentum. The hadronically decaying $W$-boson is reconstructed as a large-radius jet clustered with the Cambridge--Aachen algorithm~\cite{Cambridge--Achen1,Cambridge--Achen2} using $R=0.8$. Cut-based analyses are performed with MadAnalysis~5~\cite{MadAnalysis}. The EasyScan\_HEP package~\cite{EasyScan} is used to scan the relevant signal parameter space.

\begin{figure}[H]
    \centering
    % 第一行
    \includegraphics[width=0.4\textwidth]{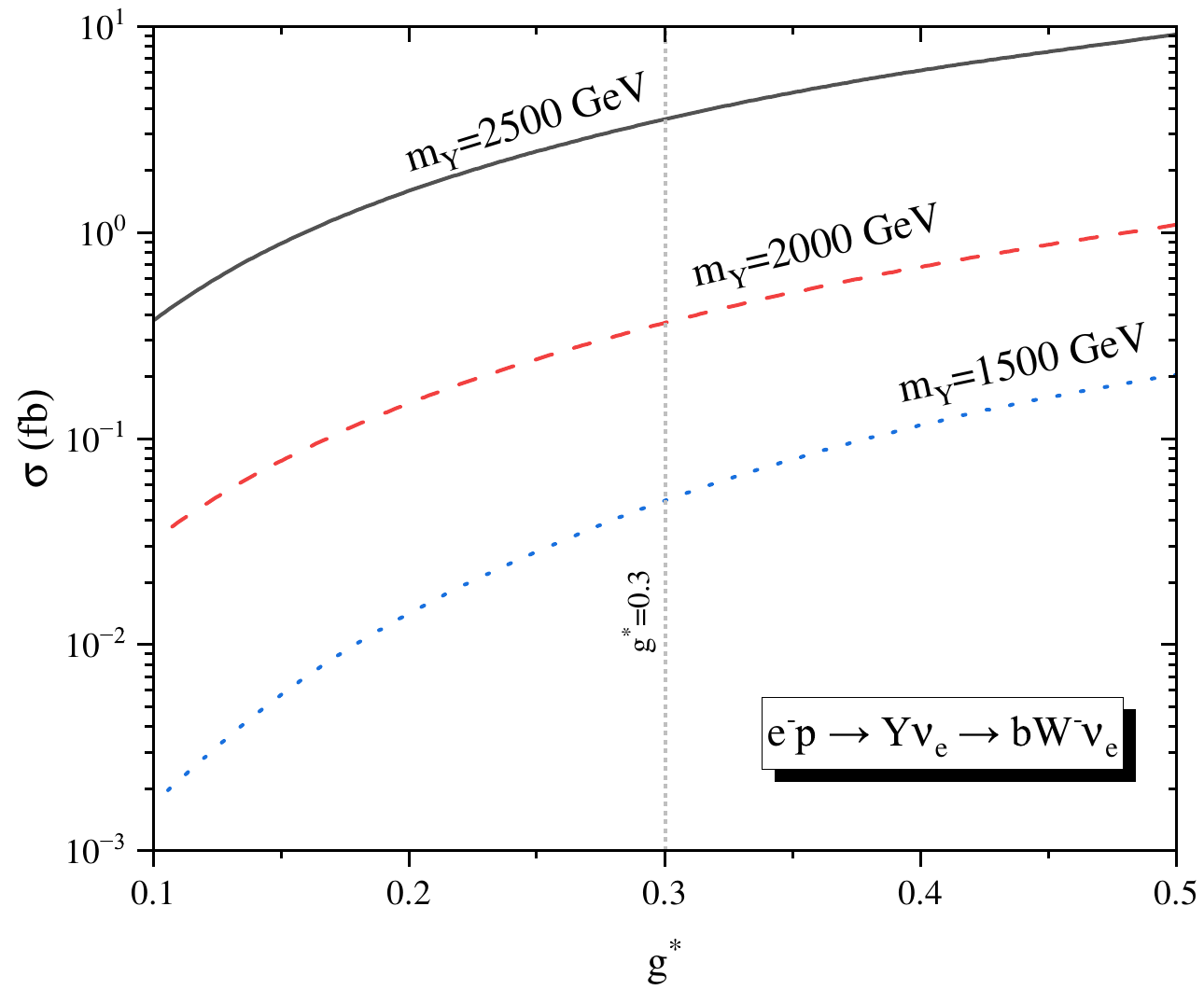}%
    \hspace{0.1\textwidth}%
    \hspace{-1cm}%
    \includegraphics[width=0.4\textwidth]{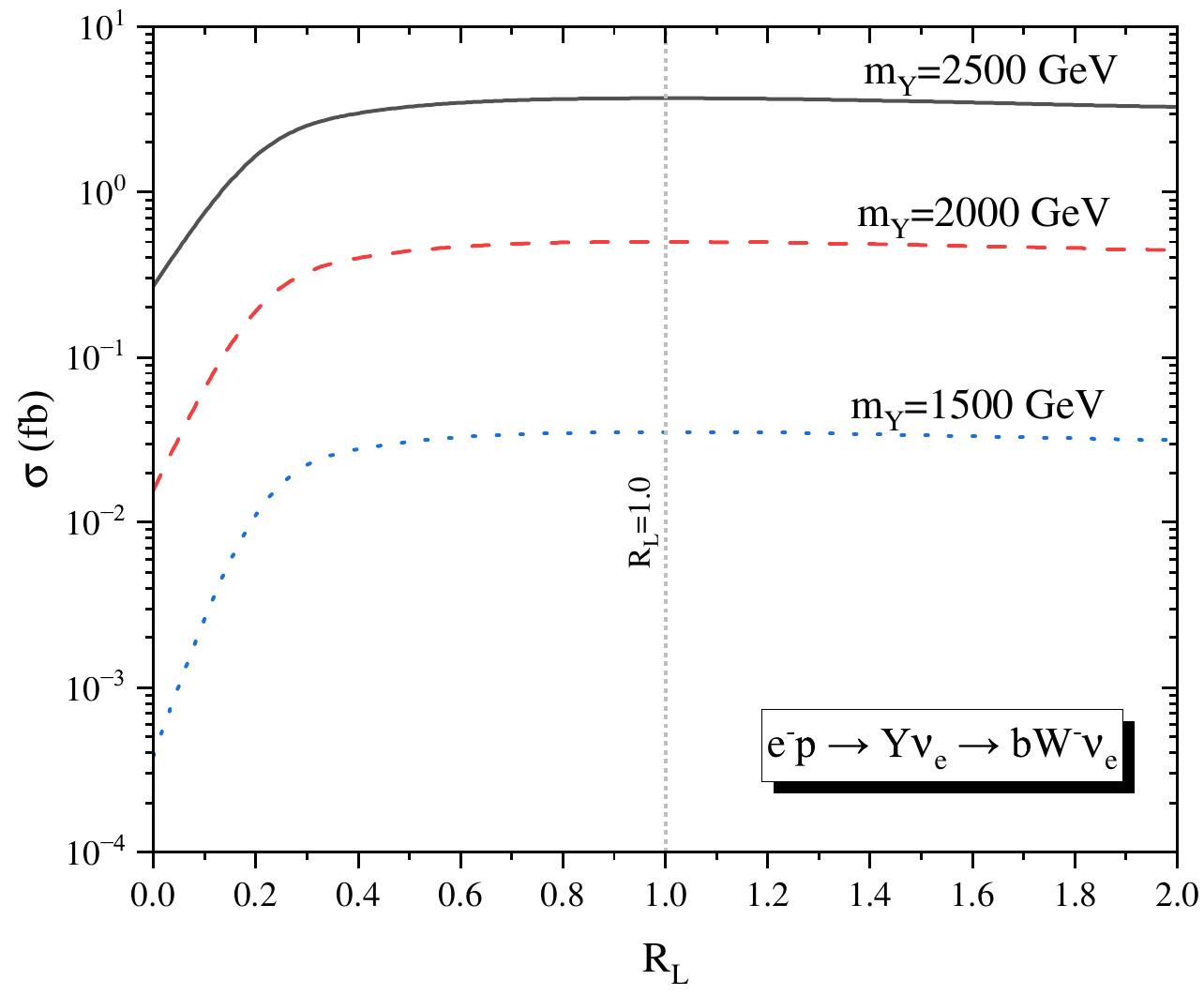}
    
   % \vspace{0.2cm}   % 两行之间的垂直间距（可按需调整）
    
    % 第二行
    %\includegraphics[width=0.35\textwidth]{ep_xsection.pdf}%
    %\hspace{0.1\textwidth}%
    %\includegraphics[width=0.35\textwidth]{antiep_xsection.pdf}
    
    \caption{Signal cross sections for the process $e^-p \to Y (\to bW)\nu_e$ at the FCC-eh as functions of $g^*$ (left) and $R_L$ (right) at $\sqrt{s}=3.46~\mathrm{TeV}$ for three benchmark masses, $m_Y=1500$, $2000$, and $2500~\mathrm{GeV}$.}
    \label{fig2}
\end{figure}

For fixed parameters $g^*=0.2$ and $R_L=0.5$, we choose three signal benchmark points:  $m_Y=1500~\mathrm{GeV}$ ($Y_{1500}$), $m_Y=2000~\mathrm{GeV}$ ($Y_{2000}$) and $m_Y=2500~\mathrm{GeV}$ ($Y_{2500}$).
We show the dependence of the signal cross section on the model parameters in Fig.~\ref{fig2}. From the left panel, we can see that the cross section increase with the coupling strength $g^*$ for fixed $R_L$. From the right panel, we can see that the signal cross section increases with $R_L$ and reaches its maximum at $R_L=1$ for fixed $g^*$. For both fixed $g^*$ and $R_L$, the cross section decreases with increasing $m_Y$ due to the phase-space suppression.

%%%%%%%%%%%%%%%%%%%%%%%%%%%%%%%%%%%%%%%%
%%%%%%%%%%%%%%%%%%%%%%%%%%%%%%%%%%%%%%%%%%%%

Since a left-handed polarized electron-beam can enhance the charged-current weak processes, we take $P_e=-80\%$~\cite{Abada2019FCC,Bruning2022}. The production cross section scales approximately as $(\sigma_{\rm pol}/\sigma_{\rm unpol})\simeq 1-P_e$, so the choice $P_e=-80\%$ enhances the signal rate by about a factor of 1.8.

\section{Event analysis}
\label{section3}

\subsection{Case~1}

\begin{figure}[htbp]
    \centering
    % 第一行
    \includegraphics[width=0.45\textwidth]{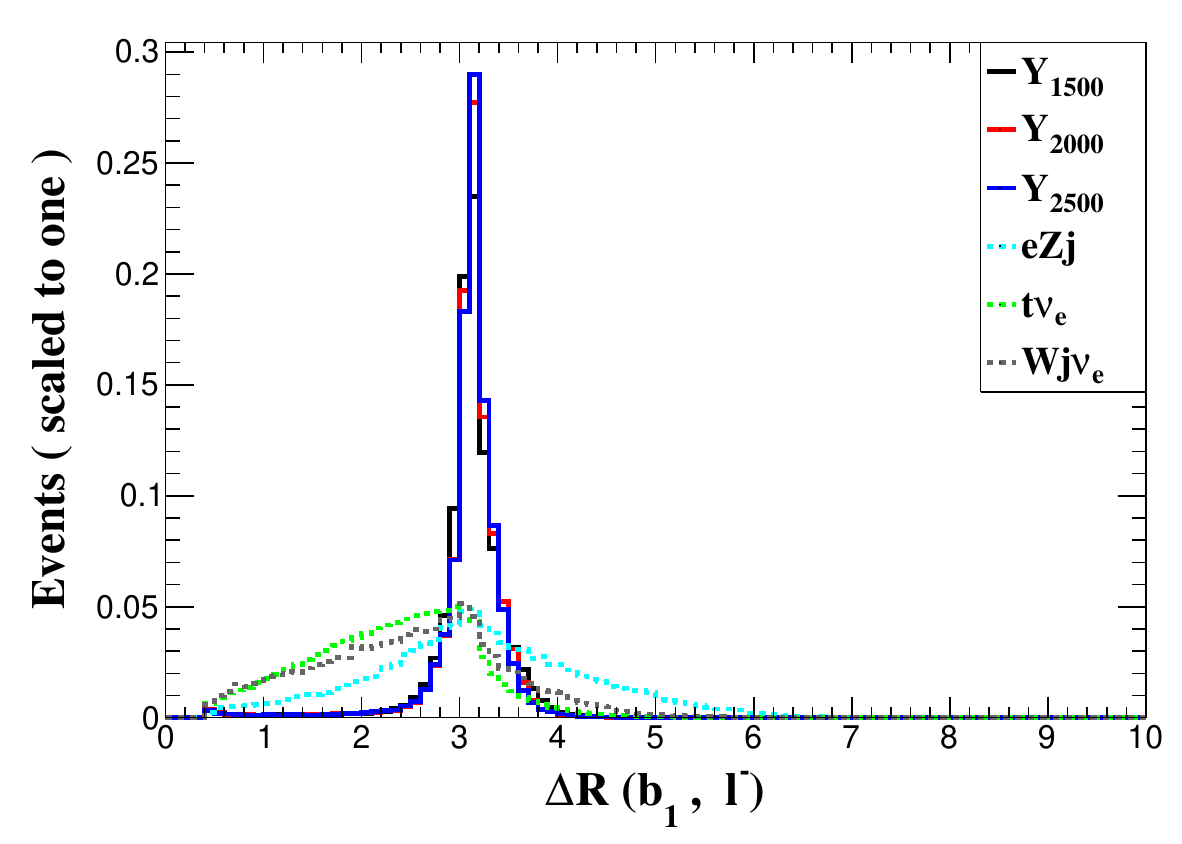}%
    \hspace{0.1\textwidth}%
    \hspace{-1cm}%
    \includegraphics[width=0.45\textwidth]{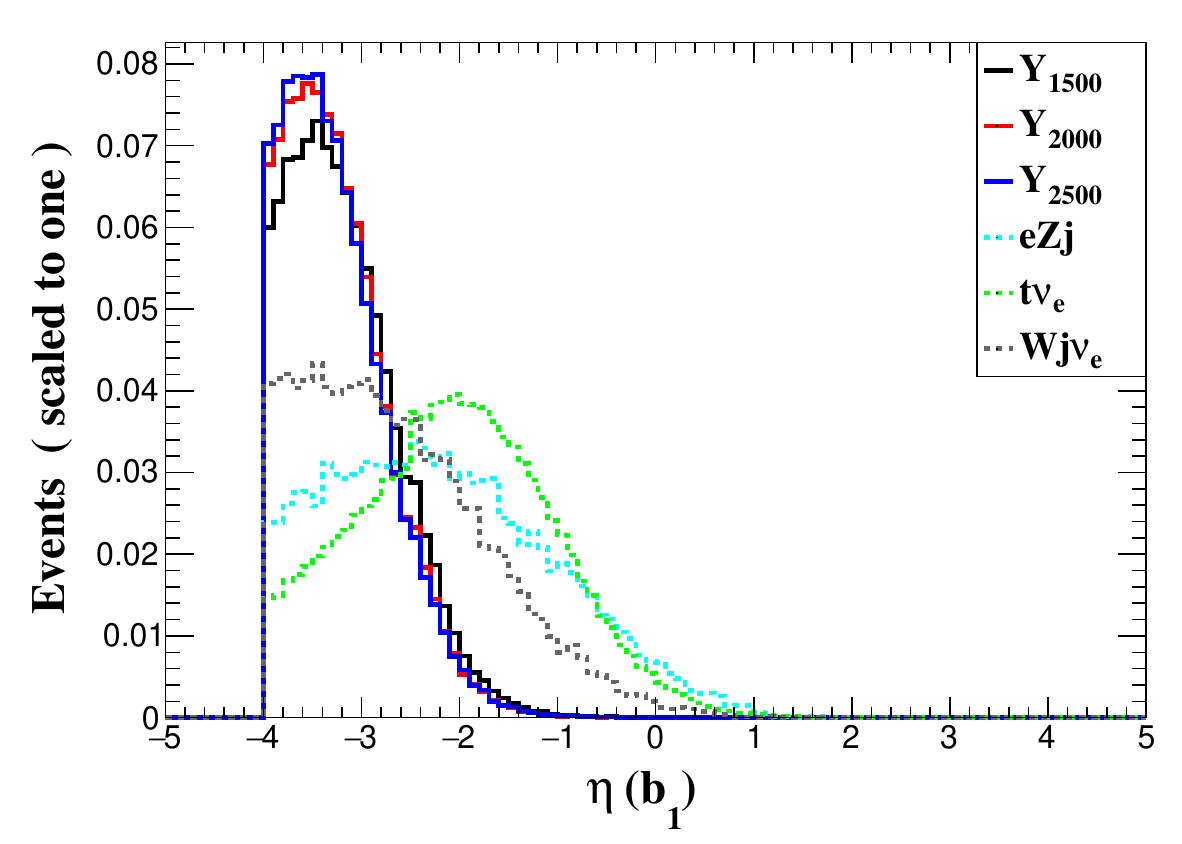}
    
    \vspace{0.2cm}   % 两行之间的垂直间距（可按需调整）
    
    % 第二行
    \includegraphics[width=0.45\textwidth]{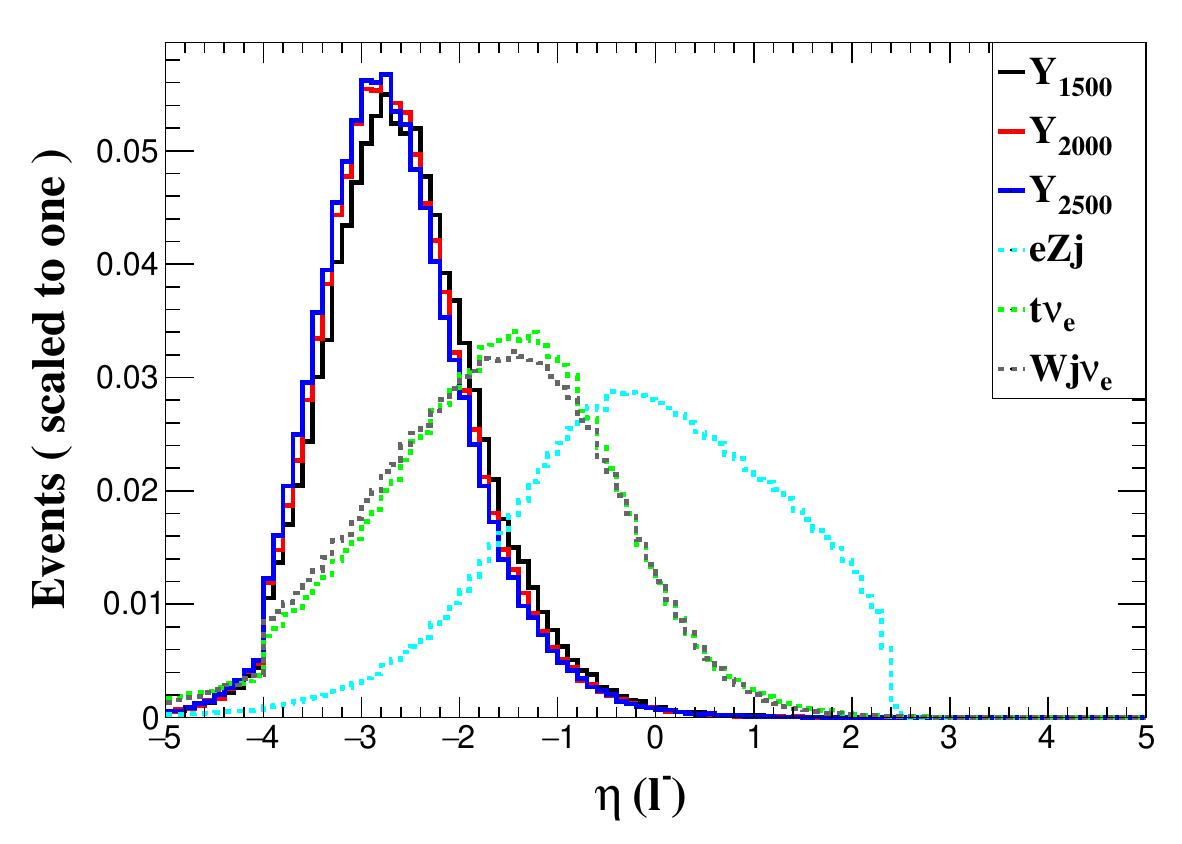}%
    \hspace{0.1\textwidth}%
    \hspace{-1cm}%
    \includegraphics[width=0.45\textwidth]{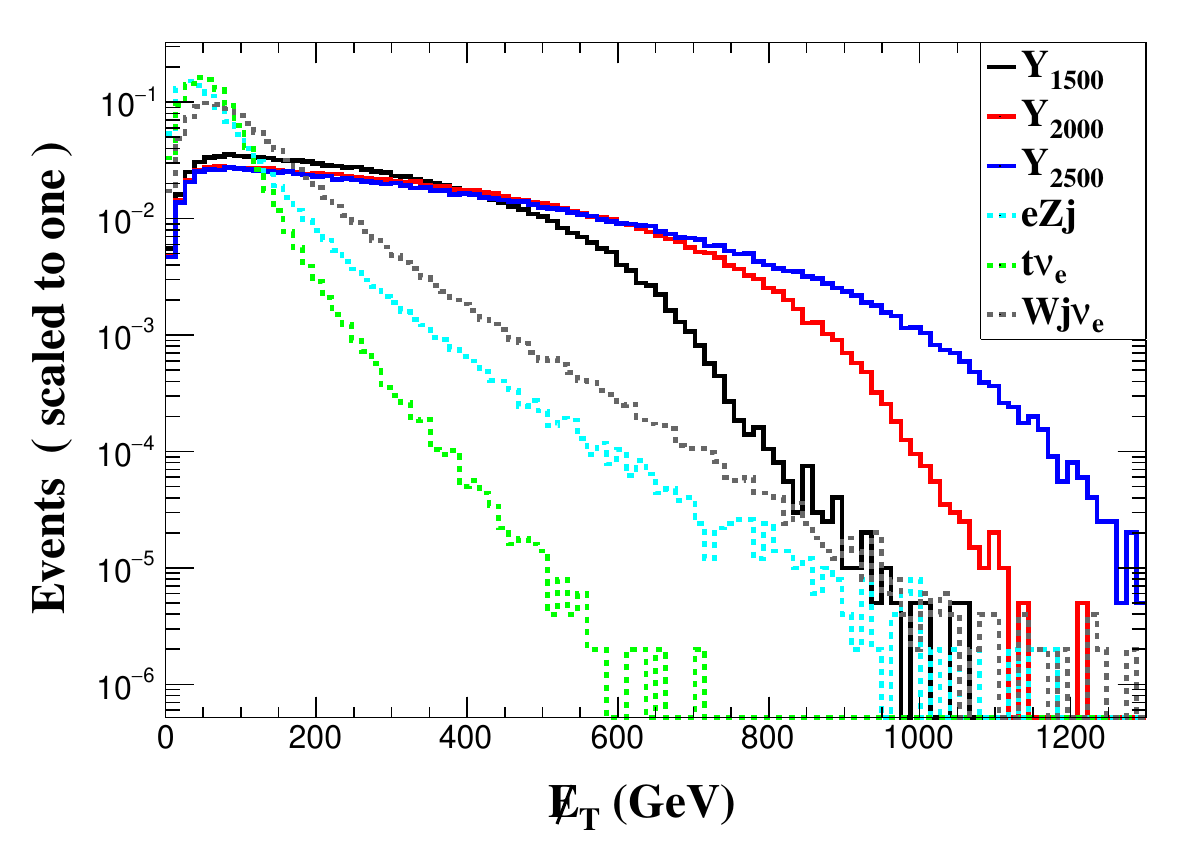}

    \vspace{0.2cm}   % 两行之间的垂直间距（可按需调整）

     % 第三行
    \includegraphics[width=0.45\textwidth]{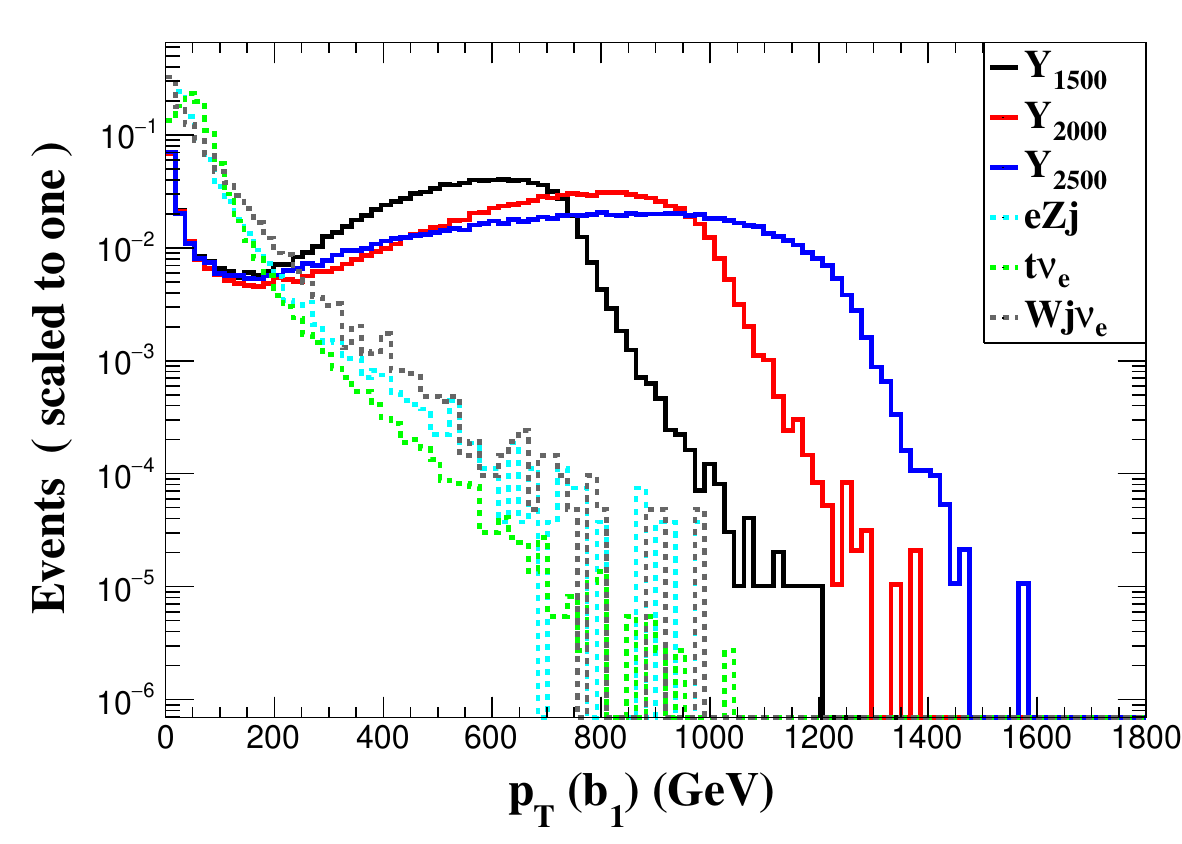}%
    \hspace{0.1\textwidth}%
    \hspace{-1cm}%
    \includegraphics[width=0.45\textwidth]{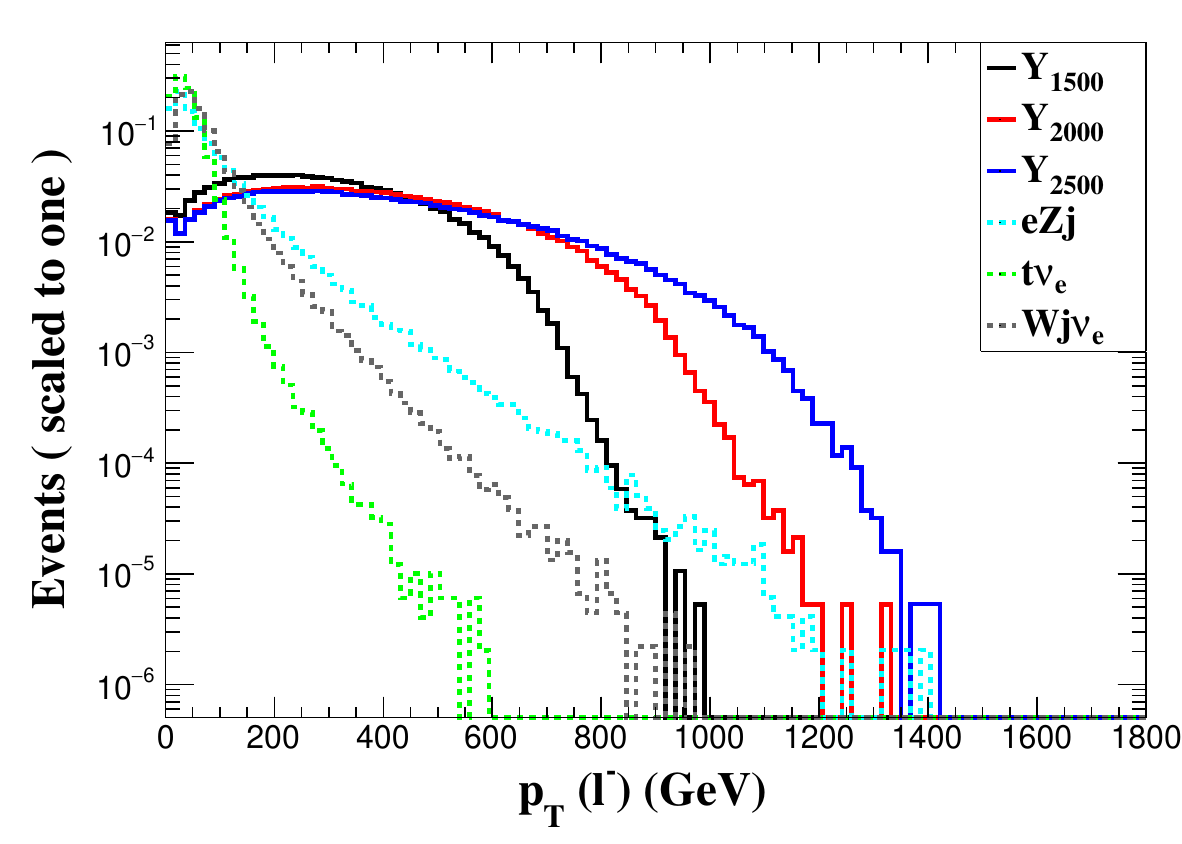}

    \vspace{0.2cm}   % 两行之间的垂直间距（可按需调整）

    % 第四行
    \includegraphics[width=0.45\textwidth]{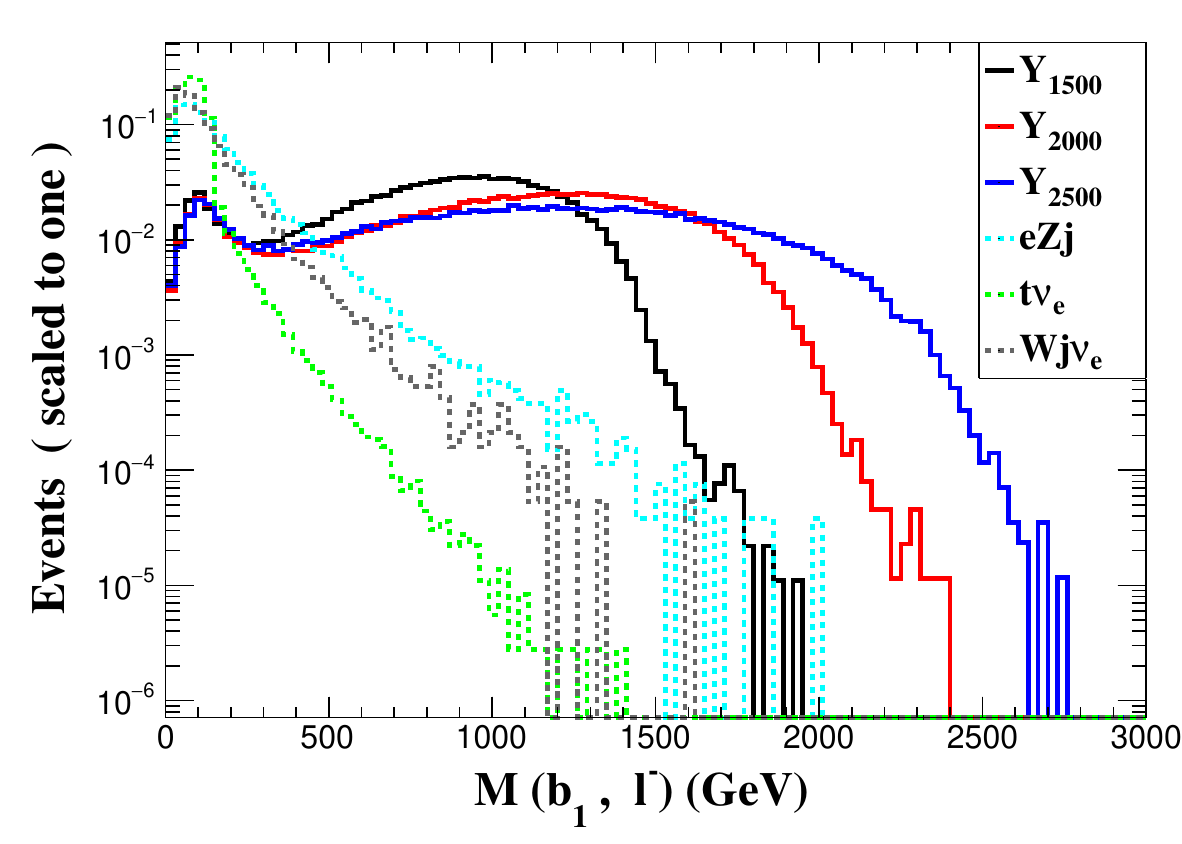}

    \caption{Normalized kinematic distributions for the signal benchmark with $m_Y=1500$, $2000$ and $2500~\mathrm{GeV}$, and the SM backgrounds at the FCC-eh with $\sqrt{s}=3.46~\mathrm{TeV}$ for Case~1.}
    \label{fig4}
\end{figure}

In Fig.~\ref{fig4}, we show the normalized kinematic distributions for the signals and the various backgrounds at the FCC-eh with $\sqrt{s}=3.46~\mathrm{TeV}$ for Case~1. The distributions at $\sqrt{s}=5.29$ and $6.9~\mathrm{TeV}$ have the similar behavior as that at $\sqrt{s}=3.46~\mathrm{TeV}$, so we only show the case at $\sqrt{s}=3.46~\mathrm{TeV}$ for example.

%%%%%%%%%%%%%%%%%%%%%%%%%%%%%%%%%%%%%%%%
%%%%%%%%%%%%%%%%%%%%%%%%%%%%%%%%%%%%%%%%%%%%

Because the $Y$ quark is TeV-scale, the $W$-boson and the $b$ quark are typically produced with large momenta. Consequently, the leading $b$-jet (denoted by $b_1$) \footnote{The subscript on the particle symbol is assigned according to the magnitude of the particle transverse momentum. For example, in the case of $b$-jets, $p_{\mathrm{T}}(b_1) > p_{\mathrm{T}}(b_2)$.} and the charged lepton from the signal processes have larger transverse momentum $p_T$ than those in the backgrounds.

The incoming electron and proton beams are taken to be along the $+z$ and $-z$ directions, respectively. The production of a heavy $Y$ quark requires partons with relatively large momentum fractions from the proton beam, so the signal system is typically boosted along the $-z$ direction. From the definition of pseudorapidity $\eta=\frac{1}{2}\ln\left(\frac{|\vec{p}|+p_z}{|\vec{p}|-p_z}\right)$, $p_z<0$ corresponds to negative pseudorapidity. Therefore, the leading $b$-jet and the charged lepton are expected to peak in the negative-$\eta$ region.

In Case~1, the presence of the neutrino prevents a direct reconstruction of the full $Y$-quark invariant mass. Therefore, we use invariant mass $M(b_1,\ell^- )$ as an effective discriminating variable for the signal. Here $M(b_1,\ell^-) = \sqrt{ (E_{b_1} + E_{\ell^-})^2 - |\vec{p}_{b_1} + \vec{p}_{\ell^-}|^2}$. Since the $b$-jet and the charged lepton are both emitted close to the $-z$ direction, their angles with respect to the neutrino can be approximated by a common angle $\theta$. Neglecting the charged-lepton and neutrino masses, one obtains:
\begin{equation}
m_Y^2 \simeq M^2(b_1,\ell^-) + 2|\vec p_{\ell^-}||\vec p_{\nu}|(1-\cos\theta).
\label{Eq10}
\end{equation}
From the transverse momentum $p_T(b_1)$ and missing transverse energy $\not\!\!E_T$ distributions in Fig.~\ref{fig4}, we can see that the typical momenta $|\vec p_{\ell^-}|$ and $|\vec p_{\nu}|$ are of the order of several hundred GeV. Therefore, the signal distribution of $M(b_1,\ell^-)$ peaks above the TeV scale.

Based on these kinematic features and distributions in Fig.~\ref{fig4}, we apply the following selection cuts to distinguish the signals from the SM backgrounds:
\begin{itemize}
\item Trigger: Exactly one negatively charged lepton and at least one $b$-jet with the particle number $N(\ell^- )=1$ and $N(b)\geq1$;
\item Cut~1: $\Delta R(b_1,\ell^- )>3$;
\item Cut~2: $\eta(\ell^- )<-2$ and $\eta(b_1)<-2$;
\item Cut~3: $p_T(b_1)>200~\mathrm{GeV}$ and $p_T(\ell^- )>100~\mathrm{GeV}$;
\item Cut~4: $M(b_1,\ell^- )>1000~\mathrm{GeV}$.
\end{itemize}

\begin{table}[htbp]
    \centering
    \caption{Cut flow of the signal and background cross sections for Case~1 at the FCC-eh with $\sqrt{s}=3.46~\mathrm{TeV}$. The benchmark parameters are fixed to $g^*=0.2$ and $R_L=0.5$.}
    \vspace{0.8cm}
    \begin{tabular}{l c c c c c c c}
    \toprule[1pt]
    \multirow{2}{*}{Cuts} & \multicolumn{3}{c}{Signals (fb)} && \multicolumn{3}{c}{Backgrounds (fb)}  \\ \cline{2-4} \cline{6-8}
    & $Y_{1500}$ & $Y_{2000}$ & $Y_{2500}$ && $eZj$ & $t\nu_e$ & $Wj\nu_e$ \\ \cline{1-8} \midrule[1pt]
    Trigger   & 2.20 & 0.291 & 0.0285 && 8.22 & 3888.8 & 54.26 \\
    Cut 1     & 1.71 & 0.239 & 0.0234 && 4.80 & 817.5 & 17.45  \\
    Cut 2     & 1.44 & 0.222 & 0.0218 && 0.0843 & 99.95 & 1.83  \\
    Cut 3     & 1.288 & 0.202 & 0.0197 && 0.0168 & 5.67 & 0.201 \\
    Cut 4     & 0.595 & 0.151 & 0.0153 && 0.0109 & 0.165 & 0.0335 \\
    Efficiency   & 11.9\% & 22.0\% & 22.1\% && 0.0066\% & 0.0026\% & 0.0022\%  \\
    \bottomrule[1pt]
    \end{tabular}  
    \label{table1}
\end{table}

When $E_p=50~\mathrm{TeV}$ is fixed, the centre-of-mass energy $\sqrt{s}$ can be enhanced by increasing the electron-beam energy. Under these conditions, producing a $Y$ quark with the same mass requires a smaller momentum fraction from the proton beam. Consequently, the longitudinal boost of the signal system along the $-z$ direction is reduced, causing the $|\eta(\ell^- )|$ and $|\eta(b_1)|$ distributions to shift toward smaller values. Accordingly, for $\sqrt{s}=5.29$ and $6.9~\mathrm{TeV}$, the Cut~2 is modified to $\eta(\ell^- )<-1$ and $\eta(b_1)<-2$. Note that we have checked the single-boson backgrounds $bW\nu_e$, $tZ\nu_\ell$ and $eZb$, as well as the diboson background $\nu_e ZWj$. After the selection cuts, their contributions are found to be negligible.

\begin{table}[htbp]
    \centering
    \caption{Cut flow of the signal and background cross sections for Case~1 at the FCC-eh with $\sqrt{s}=5.29~\mathrm{TeV}$. The benchmark parameters are fixed to $g^*=0.2$ and $R_L=0.5$.}
    \vspace{0.8cm}
    \begin{tabular}{l c c c c c c c}
    \toprule[1pt]
    \multirow{2}{*}{Cuts} & \multicolumn{3}{c}{Signals (fb)} && \multicolumn{3}{c}{Backgrounds (fb)}  \\ \cline{2-4} \cline{6-8}
    & $Y_{1500}$ & $Y_{2000}$ & $Y_{2500}$ && $eZj$ & $t\nu_e$ & $Wj\nu_e$ \\ \cline{1-8} \midrule[1pt]
    Trigger   & 14.46 & 4.95 & 1.48 && 16.54 & 8255.0 & 129.5  \\
    Cut 1     & 11.28 & 4.12 & 1.268 && 10.58 & 1896.5 & 48.73  \\
    Cut 2     & 8.76 & 3.49 & 1.125 && 0.35 & 291.8 & 8.4  \\
    Cut 3     & 7.56 & 3.11 & 1.018 && 0.0639 & 17.38 & 0.661 \\
    Cut 4     & 3.60 & 2.44 & 0.908 && 0.0399 & 1.82 & 0.175 \\
    Efficiency   & 13.6\% & 26.4\% & 32.1\% && 0.015\% & 0.014\% & 0.0062\%  \\
    \bottomrule[1pt]
    \end{tabular}
    \label{table2}
\end{table}

\begin{table}[htbp]
    \centering
    \caption{Cut flow of the signal and background cross sections for Case~1 at the FCC-eh with $\sqrt{s}=6.9~\mathrm{TeV}$. The benchmark parameters are fixed to $g^*=0.2$ and $R_L=0.5$.}
    \vspace{0.8cm}
    \begin{tabular}{l c c c c c c c}
    \toprule[1pt]
    \multirow{2}{*}{Cuts} & \multicolumn{3}{c}{Signals (fb)} && \multicolumn{3}{c}{Backgrounds (fb)} \\ \cline{2-4} \cline{6-8}
    & $Y_{1500}$ & $Y_{2000}$ & $Y_{2500}$ && $eZj$ & $t\nu_e$ & $Wj\nu_e$ \\ \cline{1-8} \midrule[1pt]
    Trigger   & 30.84 & 14.24 & 6.17 && 24.85 & 12604.5 & 212.3 \\
    Cut 1     & 23.62 & 11.67 & 5.23 && 16.65 & 3027.8 & 87.79 \\
    Cut 2     & 12.99 & 7.46 & 3.66 && 0.223 & 304.3 & 9.37 \\
    Cut 3     & 11.19 & 6.69 & 3.33 && 0.027 & 17.15 & 0.648 \\
    Cut 4     & 5.4 & 5.29 & 3.01 && 0.0185 & 1.84 & 0.203 \\
    Efficiency   & 10.2\% & 21.5\% & 28.0\% && 0.0052\% & 0.0098\% & 0.005\%  \\
    \bottomrule[1pt]
    \end{tabular}
    \label{table3}
\end{table}

As shown in Tab.~\ref{table1}, after the $\eta$-based selections in Cuts~1 and~2, the dominant background $t\nu_e$ is reduced from $3888.8~\mathrm{fb}$ to $99.95~\mathrm{fb}$, while the $eZj$ and $Wj\nu_e$ backgrounds are suppressed from $8.22~\mathrm{fb}$ and $54.26~\mathrm{fb}$ to $0.0843~\mathrm{fb}$ and $1.83~\mathrm{fb}$, respectively. To further reduce the $t\nu_e$ background, a hard $p_T$ selection is imposed, which decreases $t\nu_e$ from $99.95~\mathrm{fb}$ to $5.67~\mathrm{fb}$. Finally, the invariant-mass cut $M(b_1,\ell^- )>1000~\mathrm{GeV}$ further reduces the $t\nu_e$ and $Wj\nu_e$ backgrounds from $5.67~\mathrm{fb}$ and $0.201~\mathrm{fb}$ to $0.165~\mathrm{fb}$ and $0.0335~\mathrm{fb}$, respectively. Tabs.~\ref{table1}--\ref{table3} show that the selection cuts suppress the backgrounds while maintaining a sizable signal efficiency. The full selection retains an average signal efficiency of about $20\%$ and reduces the dominant $t\nu_e$ background by four to five orders of magnitude. For example, at $\sqrt{s}=6.9~\mathrm{TeV}$, the signal efficiencies for $m_Y=1500$, $2000$, and $2500~\mathrm{GeV}$ are $10.2\%$, $21.5\%$, and $28.0\%$, respectively, while the efficiencies of the $eZj$, $t\nu_e$, and $Wj\nu_e$ backgrounds are reduced to $0.0052\%$, $0.0098\%$, and $0.0050\%$.

\subsection{Case~2}

\begin{figure}[htbp]
    \centering
    % 第一行
    \includegraphics[width=0.45\textwidth]{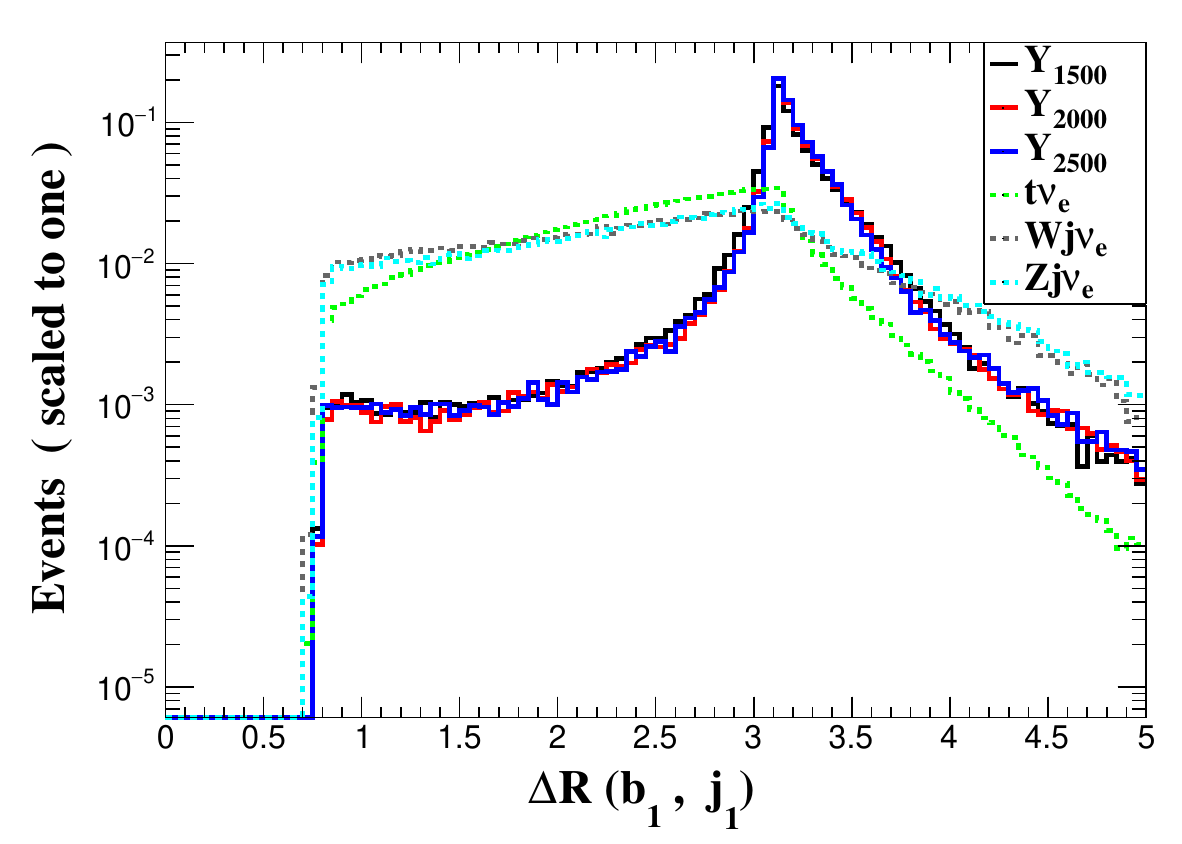}%
    \hspace{0.1\textwidth}%
    \hspace{-1cm}%
    \includegraphics[width=0.45\textwidth]{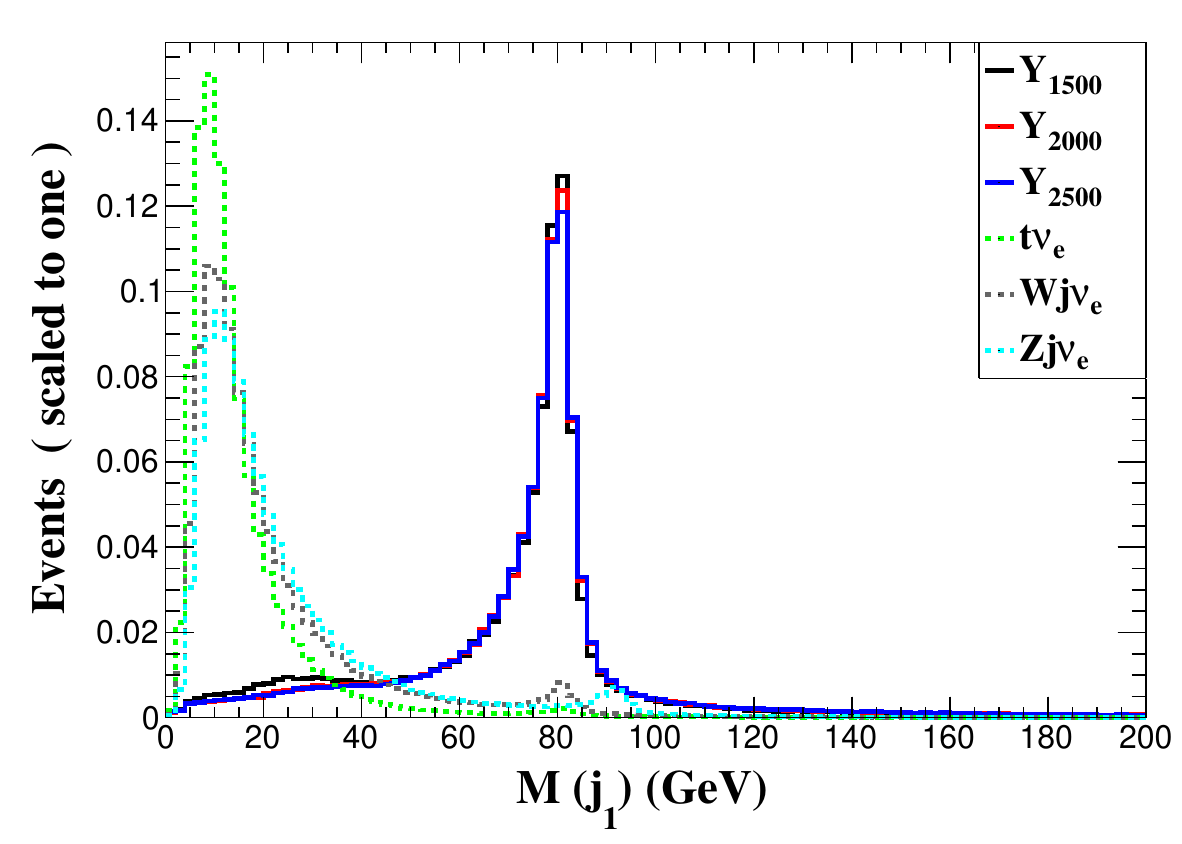}
    
    \vspace{0.2cm}   % 两行之间的垂直间距（可按需调整）
    
    % 第二行
    \includegraphics[width=0.45\textwidth]{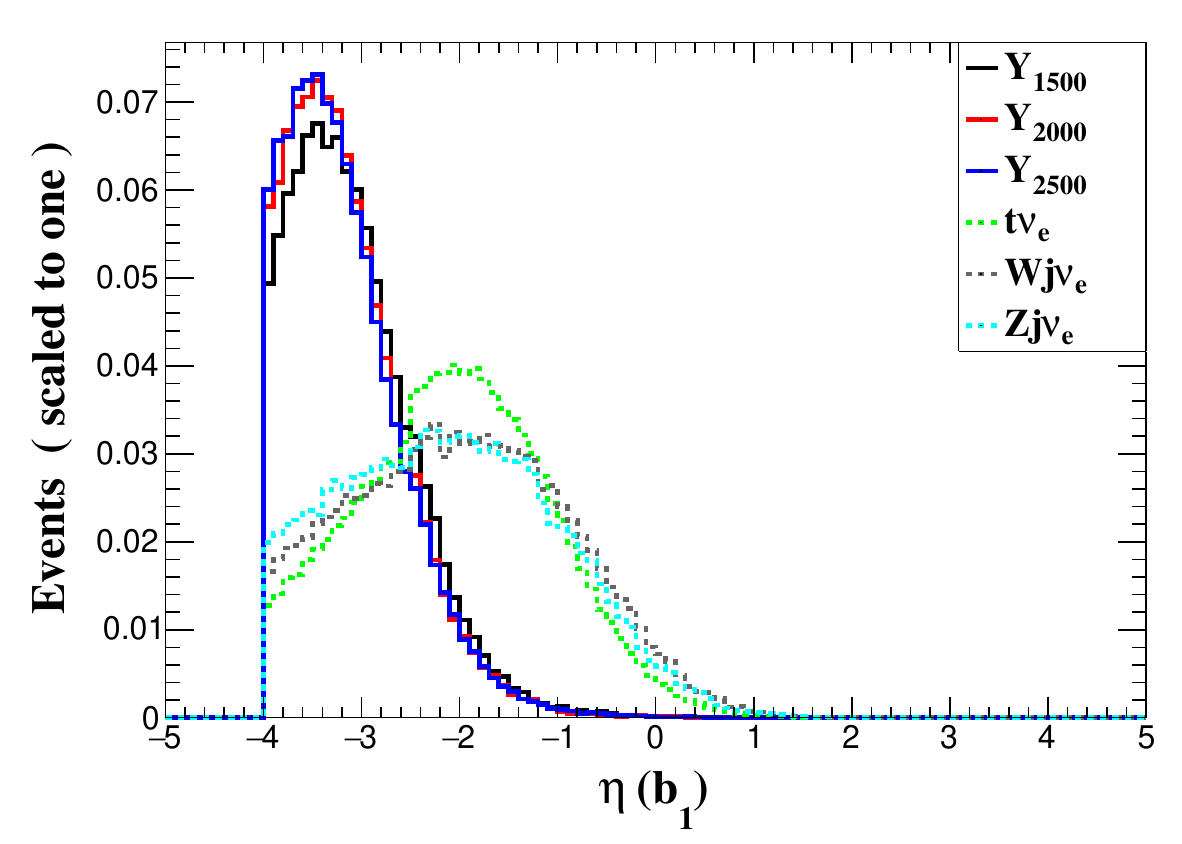}%
    \hspace{0.1\textwidth}%
    \hspace{-1cm}%
    \includegraphics[width=0.45\textwidth]{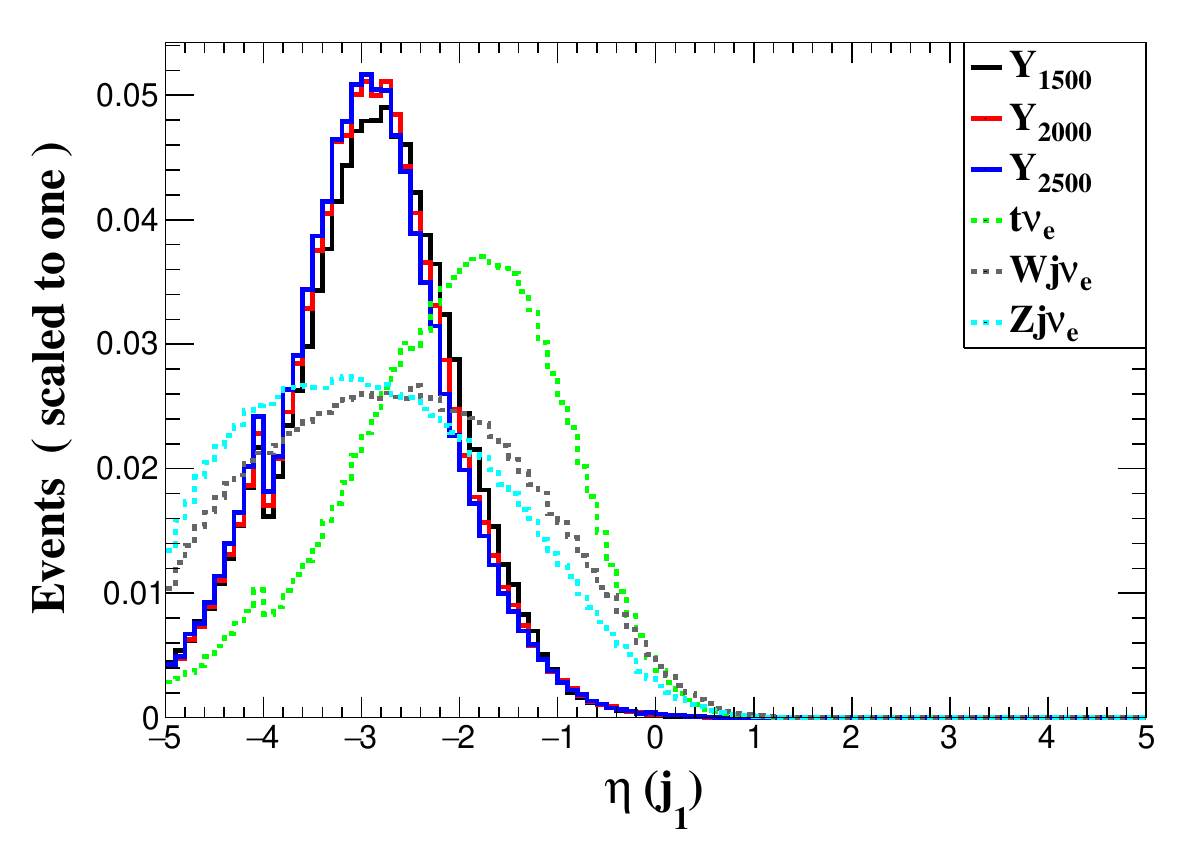}

    \vspace{0.2cm}   % 两行之间的垂直间距（可按需调整）

     % 第三行
    \includegraphics[width=0.45\textwidth]{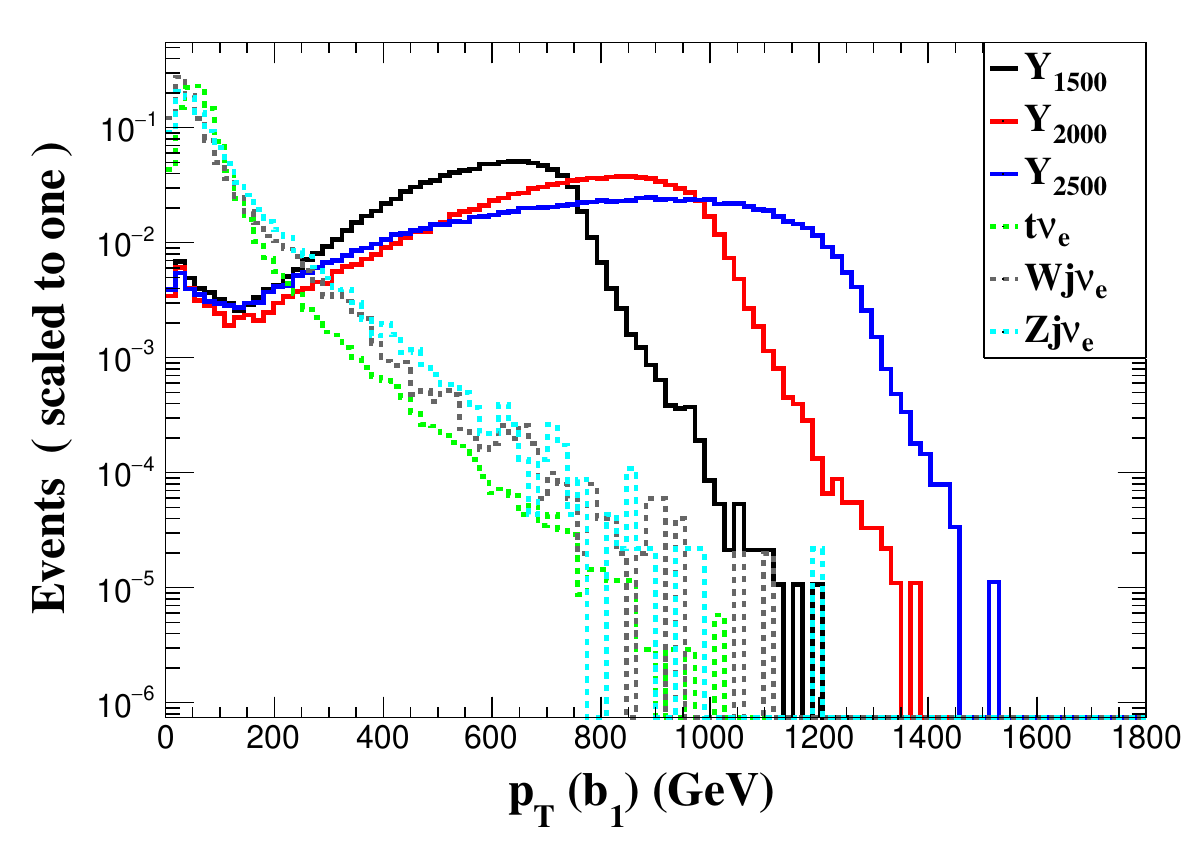}%
    \hspace{0.1\textwidth}%
    \hspace{-1cm}%
    \includegraphics[width=0.45\textwidth]{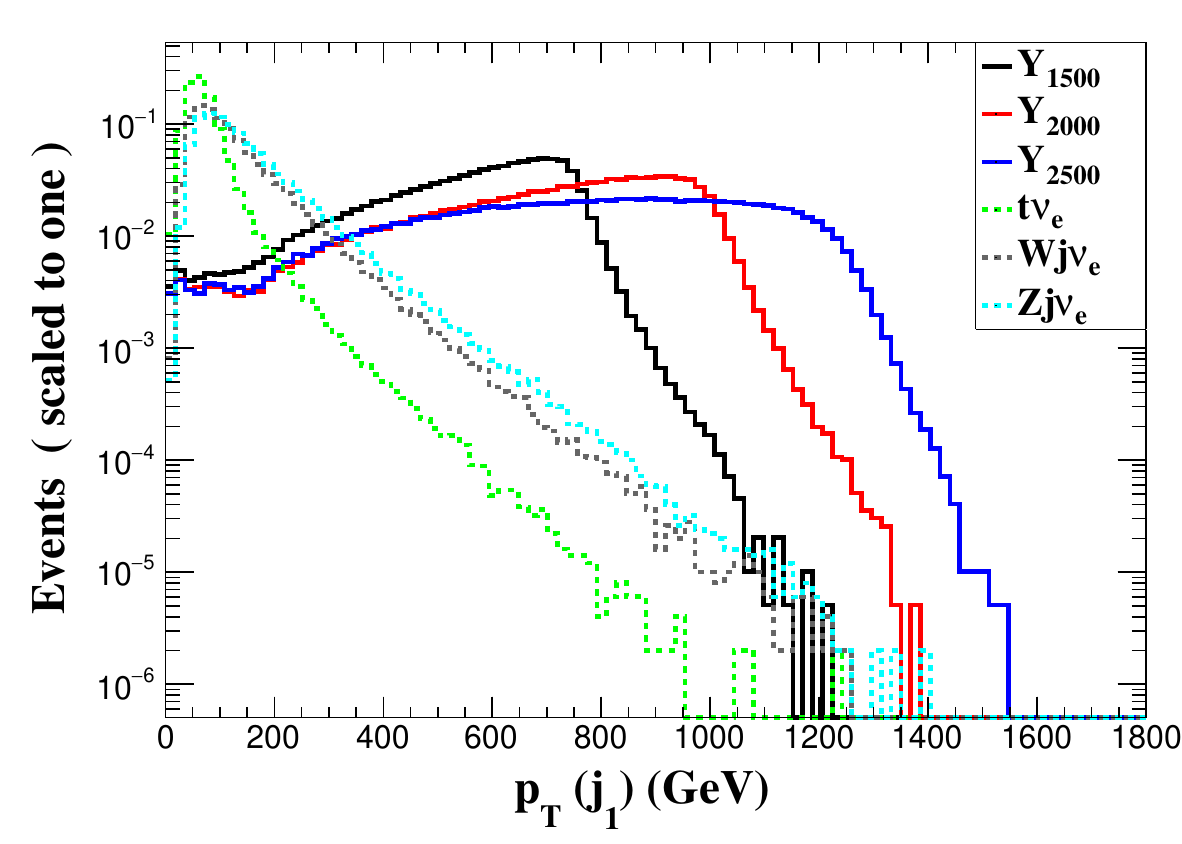}

    \vspace{0.2cm}   % 两行之间的垂直间距（可按需调整）

    % 第四行
    \includegraphics[width=0.45\textwidth]{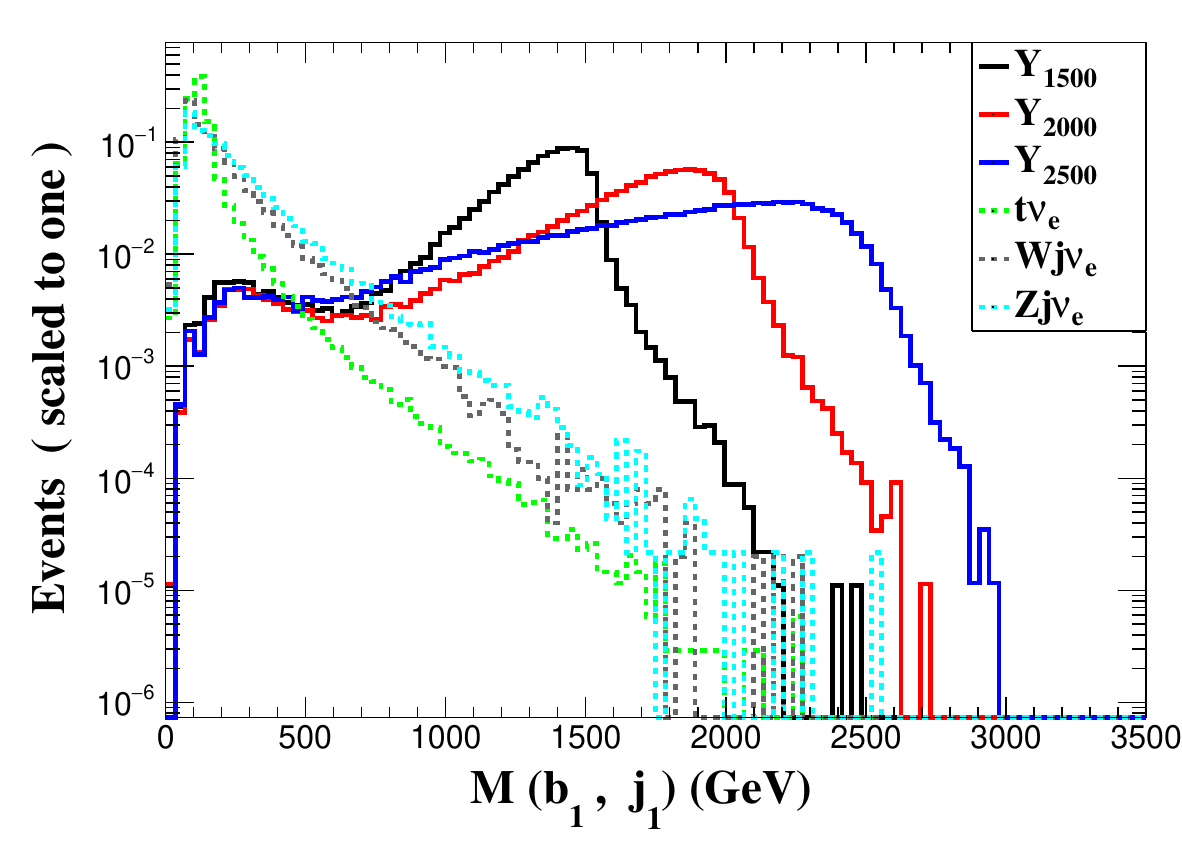}

    \caption{Normalized kinematic distributions for the signal benchmark with $m_Y=1500$, $2000$, and $2500~\mathrm{GeV}$ and the backgrounds at the FCC-eh with $\sqrt{s}=3.46~\mathrm{TeV}$ for Case~2.}

    \label{fig5}
\end{figure}

In addition to the $\eta$ and $p_T$ features discussed in Case~1, the hadronic channel is characterized by a boosted $W$-boson from the $Y$-quark decay, whose hadronic decay products are collimated with a small angular separation $\Delta R$. This feature enables the hadronically decaying $W$-boson to be reconstructed as a $W$-jet, denoted by $j_1$. In Fig.~\ref{fig5}, we show the normalized kinematic distributions of the three signal benchmark and the backgrounds for Case~2 at the FCC-eh with $\sqrt{s}=3.46~\mathrm{TeV}$. For the signal processes, the invariant-mass distribution $M(j_1)$ peaks around the $W$ mass, while $M(b_1,j_1)$ exhibits clear peaks near the benchmark masses $m_Y=1500$, $2000$ and $2500~\mathrm{GeV}$. For the backgrounds, the contamination from light jets in the $W$-jet reconstruction leads to small peaks around the $W$ mass for the $t\nu_e$ and $Wj\nu_e$ backgrounds, while the $Zj\nu_e$ background peaks around the $Z$ mass. The distributions at $\sqrt{s}=5.29$ and $6.9~\mathrm{TeV}$ show the similar qualitative behavior.

%%%%%%%%%%%%%%%%%%%%%%%%%%%%%%%%%%%%%%%%
%%%%%%%%%%%%%%%%%%%%%%%%%%%%%%%%%%%%%%%%%%%%

Based on these kinematic features and distributions in Fig.~\ref{fig5}, we apply the following selection cuts:
\begin{itemize}
\item Trigger: At least one b-jet with $N(b)\geq1$;
\item Cut~1: $\Delta R(b_1,j_1)>3$;
\item Cut~2: $\eta(b_1)<-2$ and $\eta(j_1)<-2$;
\item Cut~3: $p_T(b_1)>200~\mathrm{GeV}$ and $p_T(j_1)>300~\mathrm{GeV}$;
\item Cut~4: $70~\mathrm{GeV}<M(j_1)<90~\mathrm{GeV}$;
\item Cut~5: $M(b_1,j_1)>1400~\mathrm{GeV}$.
\end{itemize}

\begin{table}[htbp]
    \centering
    \caption{Cut flow of the signal and background cross sections for Case~2 at the FCC-eh with $\sqrt{s}=3.46~\mathrm{TeV}$. The benchmark parameters are fixed to $g^*=0.2$ and $R_L=0.5$.}
    \vspace{0.8cm}
    \begin{tabular}{l c c c c c c c}
    \toprule[1pt]
    \multirow{2}{*}{Cuts} & \multicolumn{3}{c}{Signals (fb)} && \multicolumn{3}{c}{Backgrounds (fb)}  \\ \cline{2-4} \cline{6-8}
    & $Y_{1500}$ & $Y_{2000}$ & $Y_{2500}$ && $t\nu_e$ & $Wj\nu_e$ & $Zj\nu_e$ \\ \cline{1-8} \midrule[1pt]
    Trigger   & 7.05 & 0.936 & 0.0926 && 13146.7 & 458.3 & 188.1  \\
    Cut 1     & 5.91 & 0.8 & 0.0797 && 3066.5 & 142.3 & 64.24 \\
    Cut 2     & 5.11 & 0.744 & 0.0737 && 691.0 & 26.54 & 13.1 \\
    Cut 3     & 4.7 & 0.716 & 0.0698 && 14.67 & 0.557 & 0.451  \\
    Cut 4     & 3.18 & 0.46 & 0.044 && 2.28 & 0.137 & 0.115 \\
    Cut 5     & 1.6 & 0.426 & 0.0389 && 0.114 & 0.0274 & 0.0205 \\
    Efficiency   & 10.6\% & 20.7\% & 18.7\% && 0.0006\% & 0.0006\% & 0.001\% \\
    \bottomrule[1pt]
    \end{tabular}
    \label{table4}
\end{table}

Similar to Case~1, for $\sqrt{s}=5.29$ and $6.9~\mathrm{TeV}$ in Case~2, Cut~2 is modified to $\eta(b_1) < -2$ and $\eta(j_1) < -1$.  We have also checked the single-boson backgrounds $bW\nu_e$ and $tZ\nu_\ell$, as well as the diboson background $\nu_e ZWj$. After the selection cuts, their contributions are found to be negligible. 

\begin{table}[htbp]
    \centering
    \caption{Cut flow of the signal and background cross sections for Case~2 at the FCC-eh with $\sqrt{s}=5.29~\mathrm{TeV}$. The benchmark parameters are fixed to $g^*=0.2$ and $R_L=0.5$.}
    \vspace{0.8cm}
    \begin{tabular}{l c c c c c c c}
    \toprule[1pt]
    \multirow{2}{*}{Cuts} & \multicolumn{3}{c}{Signals (fb)} && \multicolumn{3}{c}{Backgrounds (fb)}  \\ \cline{2-4} \cline{6-8}
    & $Y_{1500}$ & $Y_{2000}$ & $Y_{2500}$ && $t\nu_e$ & $Wj\nu_e$ & $Zj\nu_e$ \\ \cline{1-8} \midrule[1pt]
    Trigger   & 44.79 & 15.38 & 4.64 && 28048.6 & 933.7 & 385.9 \\
    Cut 1     & 36.39 & 13.00 & 4.02 && 6993.2 & 325.7 & 144.1 \\
    Cut 2     & 28.17 & 10.86 & 3.50 && 1461.2 & 47.63 & 23.39 \\
    Cut 3     & 24.96 & 10.40 & 3.41 && 48.06 & 1.37 & 0.752 \\
    Cut 4     & 17.64 & 7.1 & 2.28 && 7.35 & 0.559 & 0.188 \\
    Cut 5     & 9.83 & 6.85 & 2.25 && 1.06 & 0.237 & 0.0827 \\
    Efficiency   & 12.4\% & 24.7\% & 26.5\% && 0.0028\% & 0.0028\% & 0.0022\% \\
    \bottomrule[1pt]
    \end{tabular}
    \label{table5}
\end{table}

\begin{table}[htbp]
    \centering
    \caption{Cut flow of the signal and background cross sections for Case~2 at the FCC-eh with $\sqrt{s}=6.9~\mathrm{TeV}$. The benchmark parameters are fixed to $g^*=0.2$ and $R_L=0.5$.}
    \vspace{0.8cm}
    \begin{tabular}{l c c c c c c c}
    \toprule[1pt]
    \multirow{2}{*}{Cuts} & \multicolumn{3}{c}{Signals (fb)} && \multicolumn{3}{c}{Backgrounds (fb)} \\ \cline{2-4} \cline{6-8}
    & $Y_{1500}$ & $Y_{2000}$ & $Y_{2500}$ && $t\nu_e$ & $Wj\nu_e$ & $Zj\nu_e$ \\ \cline{1-8} \midrule[1pt]
    Trigger   & 96.08 & 44.07 & 19.12 && 42905 & 1404.4 & 577.1 \\
    Cut 1     & 76.1 & 36.61 & 16.30 && 11122.9 & 520.0 & 226.8 \\
    Cut 2     & 42.54 & 23.29 & 11.29 && 1512.3 & 54.50 & 26.26 \\
    Cut 3     & 37.62 & 22.28 & 11.00 && 57.44 & 1.19 & 0.953 \\
    Cut 4     & 26.5 & 15.37 & 7.40 && 7.79 & 0.437 & 0.257 \\
    Cut 5     & 15.22 & 14.91 & 7.34 && 1.7 & 0.194 & 0.15 \\
    Efficiency   & 9.6\% & 20.2\% & 22.7\% && 0.003\% & 0.0016\% & 0.0028\% \\
    \bottomrule[1pt]
    \end{tabular}
    \label{table6}
\end{table}

As shown in Tab.~\ref{table4}, after applying Cuts~1 and~2, the dominant background $t\nu_e$ is reduced from $13146.7~\mathrm{fb}$ to $691.0~\mathrm{fb}$, while the $Wj\nu_e$ and $Zj\nu_e$ backgrounds are suppressed from $458.3~\mathrm{fb}$ and $188.1~\mathrm{fb}$ to $26.54~\mathrm{fb}$ and $13.1~\mathrm{fb}$, respectively. A hard $p_T$ selection is then imposed to further reduce the backgrounds, yielding cross sections of $14.67~\mathrm{fb}$, $0.557~\mathrm{fb}$, and $0.451~\mathrm{fb}$ for $t\nu_e$, $Wj\nu_e$, and $Zj\nu_e$, respectively. To suppress the dominant $t\nu_e$ background further, $M(j_1)$ is required to be close to the $W$ mass, reducing the $t\nu_e$ background to $2.28~\mathrm{fb}$. Finally, the invariant-mass requirement on $M(b_1,j_1)$ exploits the heavy-resonance structure and reduces the $t\nu_e$, $Wj\nu_e$, and $Zj\nu_e$ backgrounds to $0.114~\mathrm{fb}$, $0.0274~\mathrm{fb}$, and $0.0205~\mathrm{fb}$, respectively. Tabs.~\ref{table4}--\ref{table6} show that the full selection strongly suppresses the backgrounds while retaining a sizable fraction of the signal. The dominant background $t\nu_e$ is reduced by five to six orders of magnitude, whereas the average signal efficiency remains around $20\%$. At $\sqrt{s}=6.9~\mathrm{TeV}$, the signal efficiencies for $m_Y=1500$, $2000$, and $2500~\mathrm{GeV}$ are $9.6\%$, $20.2\%$, and $22.7\%$, respectively, while the efficiencies of the $t\nu_e$, $Wj\nu_e$, and $Zj\nu_e$ backgrounds are suppressed to $0.0030\%$, $0.0016\%$, and $0.0028\%$.

\section{exclusion and discovery potential}
\label{section4}
To demonstrate the collider discovery potential corresponding to different statistical significances, we use $Z_{{disc}}$ for discovery and $Z_{{excl}}$ for exclusion, respectively. These are based on hypothesis testing, and the corresponding test statistics are constructed as follows~\cite{Cowan:2010js,Kumar:2015tna}:
\begin{equation}
Z_{{excl}} = \sqrt{2 \left[ s - b \ln \left( \frac{b + s + x}{2b} \right) - \frac{1}{\delta^2} \ln \left( \frac{b - s + x}{2b} \right) \right] - (b + s - x) \left( 1 + \frac{1}{\delta^2 b} \right)},
\label{Zexcl}
\end{equation}

\begin{equation}
Z_{{disc}} = \sqrt{2 \left[ (s + b) \ln \left( \frac{(s + b)(1 + \delta^2 b)}{b + (s + b)\delta^2 b} \right) - \frac{1}{\delta^2} \ln \left( 1 + \frac{\delta^2 s}{1 + \delta^2 b} \right) \right]},
\label{Zdisc}
\end{equation}

\begin{equation}
x = \sqrt{(s + b)^2 - \frac{4\delta^2 s b^2}{1 + \delta^2 b}}.
\end{equation}
Here, $s$ and $b$ are the numbers of signal and background events, respectively, and $\delta$ is the uncertainty that inevitably appears in the measurement of the background. The exclusion capability corresponds to $Z_{\rm excl}=2$ while the discovery potential corresponds to $Z_{\rm disc}=5$. In the completely ideal case ($\delta \to 0$), Eq.~(\ref{Zexcl}) and (\ref{Zdisc}) can be simplified as:
\begin{equation}
    \mathcal{Z}_{\mathrm{excl}} =
    \sqrt{2\left[s - b\ln\left(1+\frac{s}{b}\right)\right]}
\end{equation}
    
\begin{equation}
    \mathcal{Z}_{\mathrm{disc}} =
    \sqrt{2\left[(s+b)\ln\left(1+\frac{s}{b}\right)-s\right]}
\end{equation}

\begin{figure}[htbp]
    \centering
    % 第一行
    \includegraphics[width=0.45\textwidth]{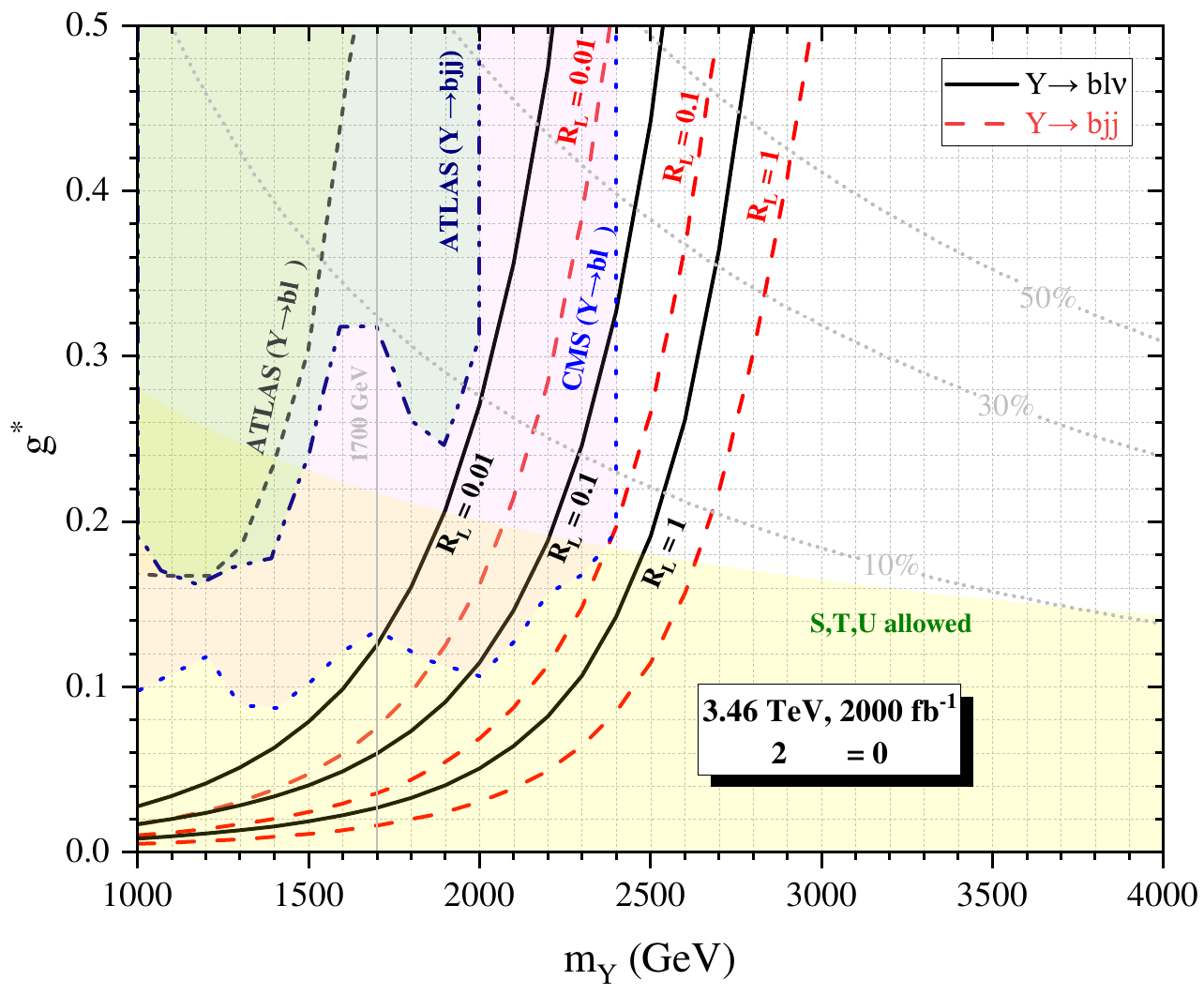}%
    \hspace{0.1\textwidth}%
    \hspace{-1cm}%
    \includegraphics[width=0.45\textwidth]{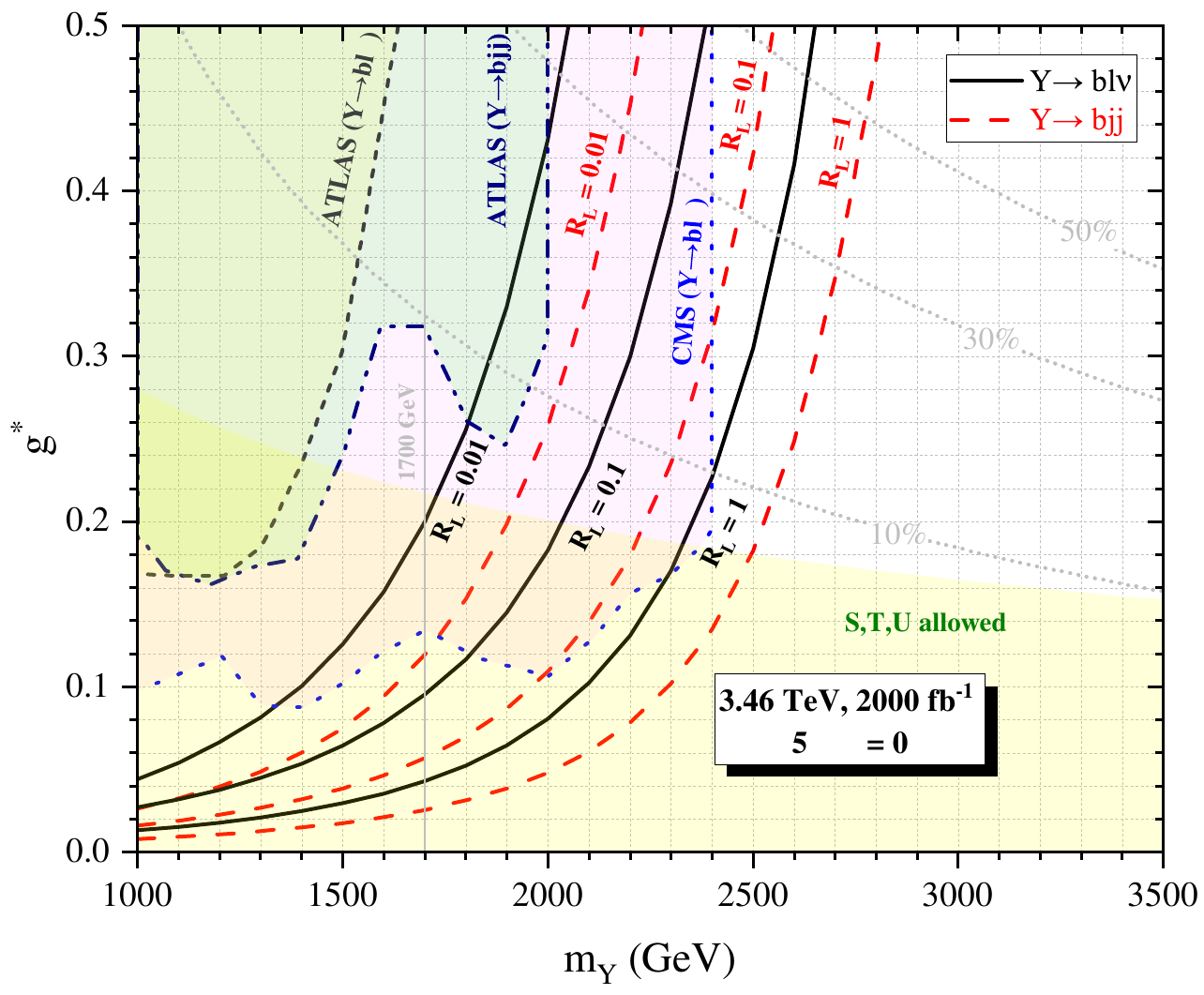}
    
    \vspace{0.2cm}   % 两行之间的垂直间距（可按需调整）
    
    % 第二行
    \includegraphics[width=0.45\textwidth]{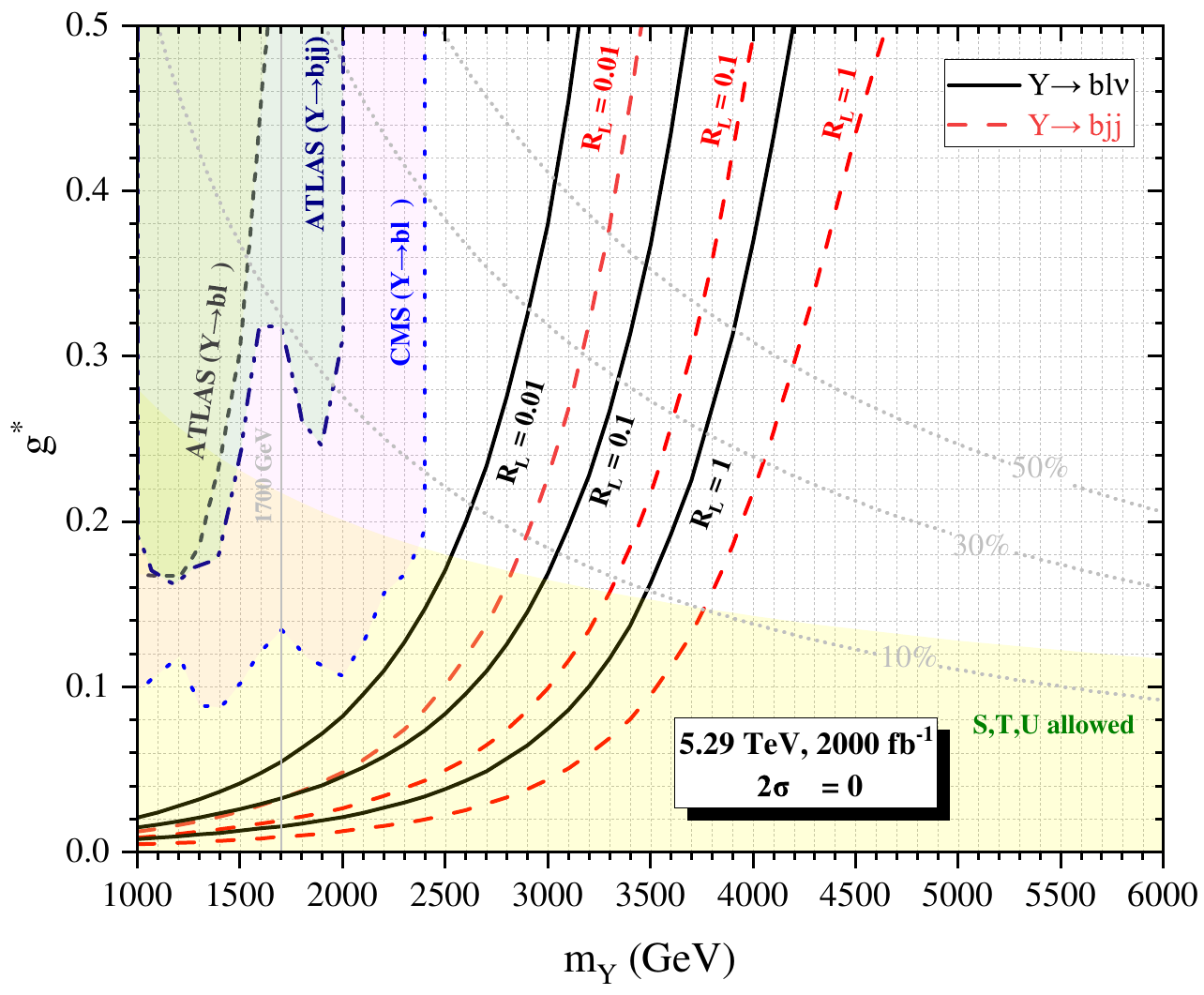}%
    \hspace{0.1\textwidth}%
    \hspace{-1cm}%
    \includegraphics[width=0.45\textwidth]{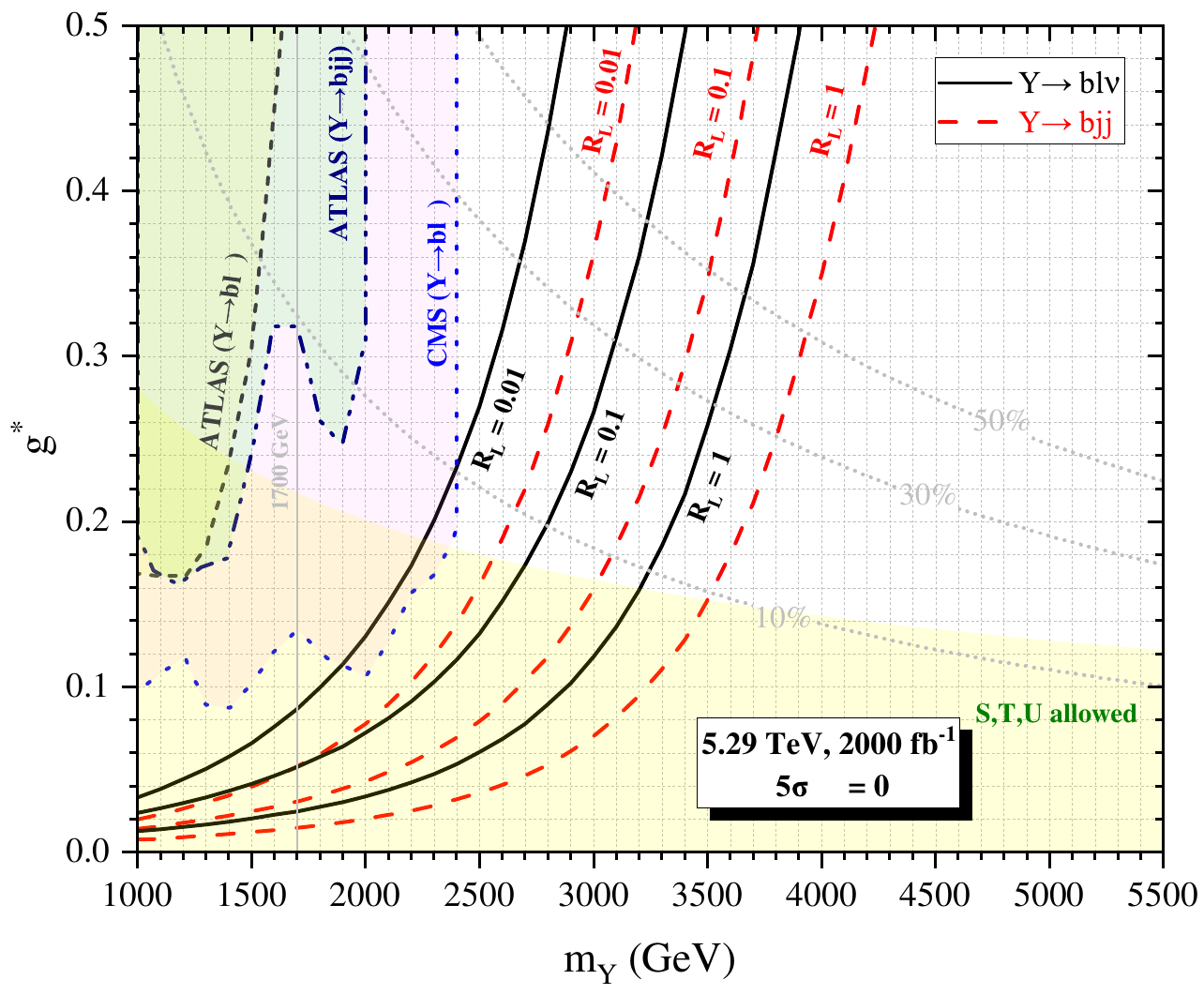}
    
    \vspace{0.2cm}

    % 第三行
    \includegraphics[width=0.45\textwidth]{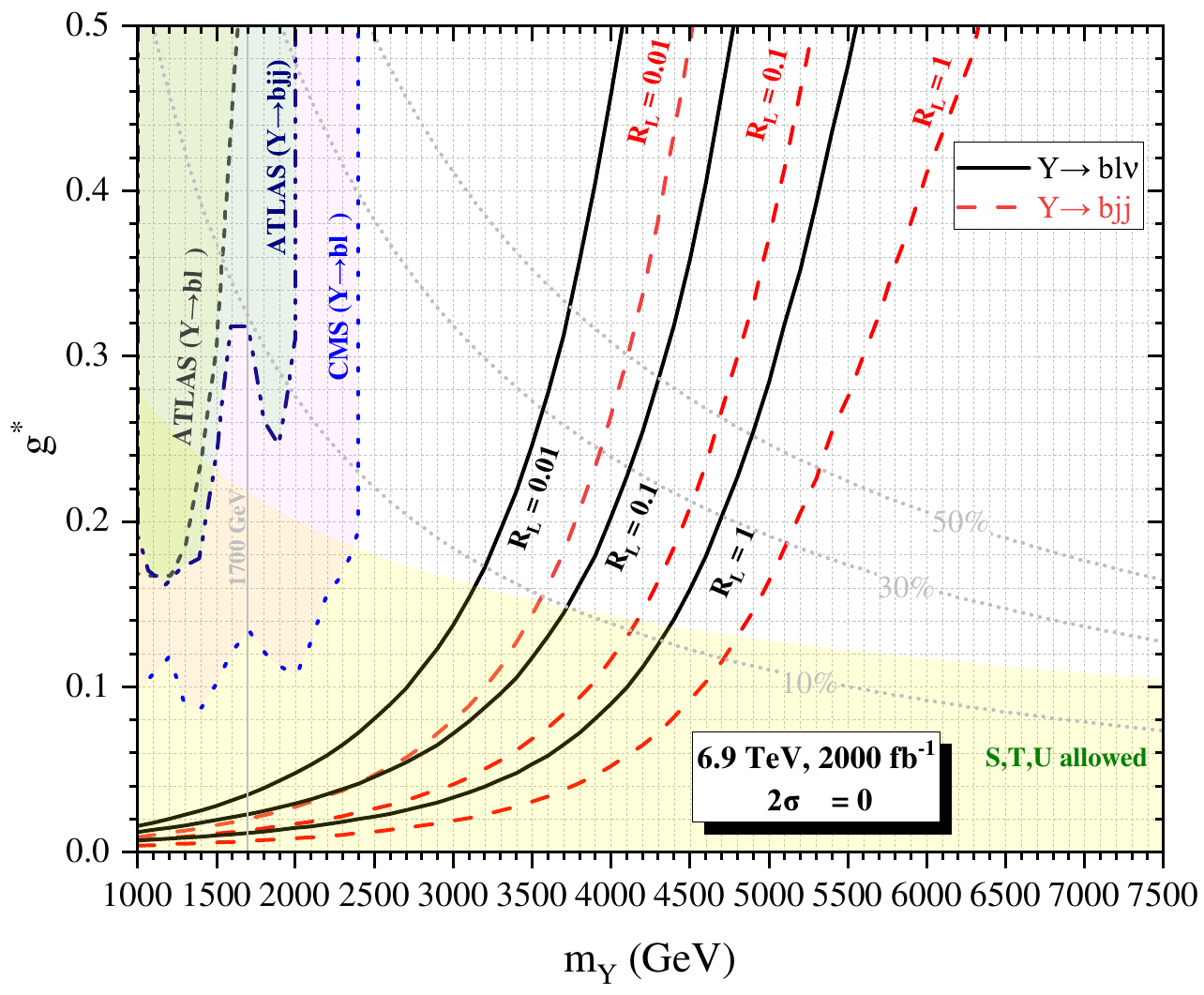}%
    \hspace{0.1\textwidth}%
    \hspace{-1cm}%
    \includegraphics[width=0.45\textwidth]{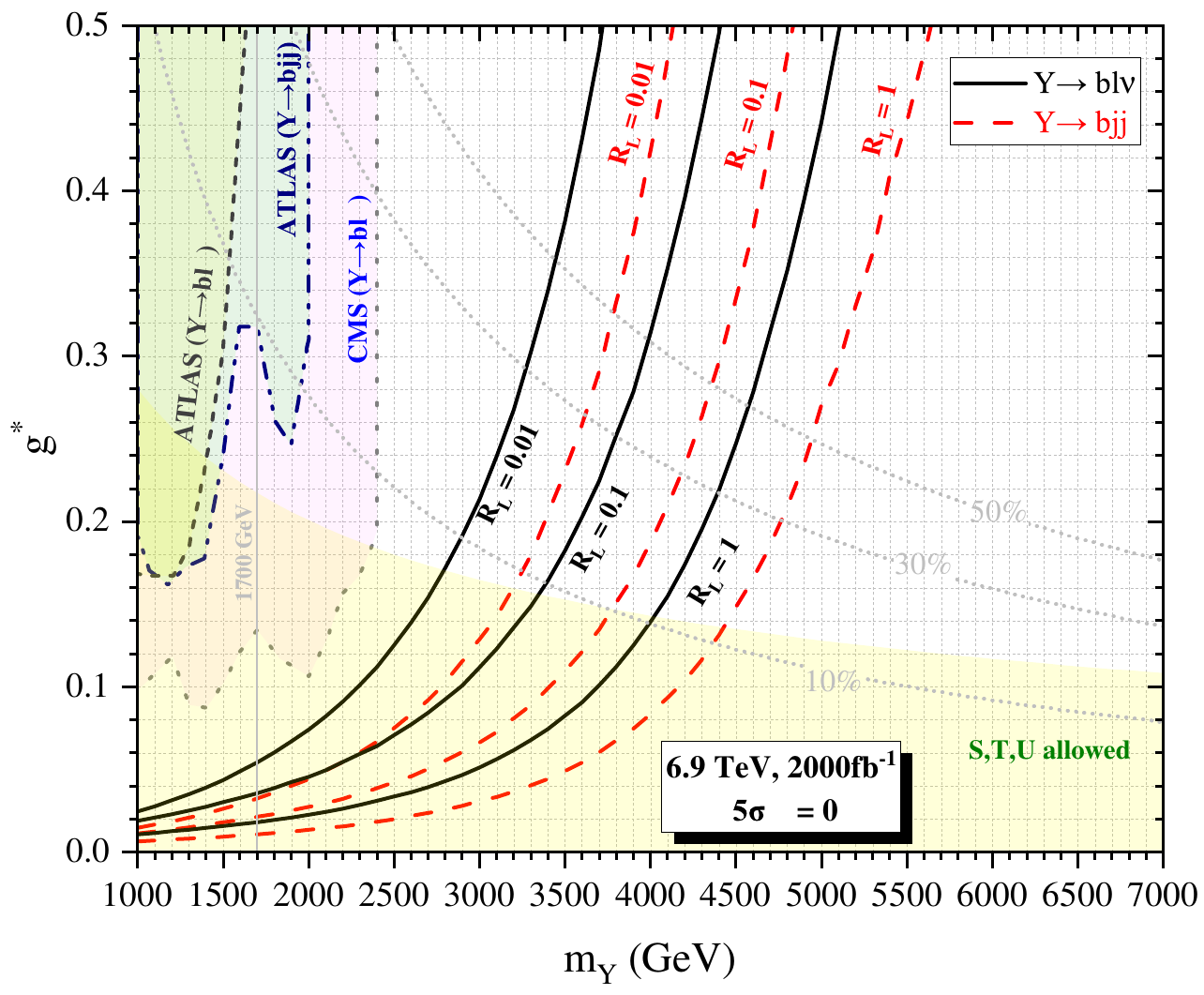}

    \caption{Contour plots of the $2\sigma$ exclusion limit (left panels) and the $5\sigma$ discovery reach (right panels) in the $g^*$--$m_Y$ plane at the FCC-eh with $\sqrt{s}=3.46$, $5.29$, and $6.9~\mathrm{TeV}$ for Case~1 and Case~2, using $\delta=0$. The solid black and dashed red contours correspond to Case~1 and Case~2, respectively. The dotted gray curves denote contours of fixed $\Gamma_Y/m_Y$.}
    \label{fig6}
\end{figure}

\begin{figure}[htbp]
    \centering
    % 第一行
    \includegraphics[width=0.45\textwidth]{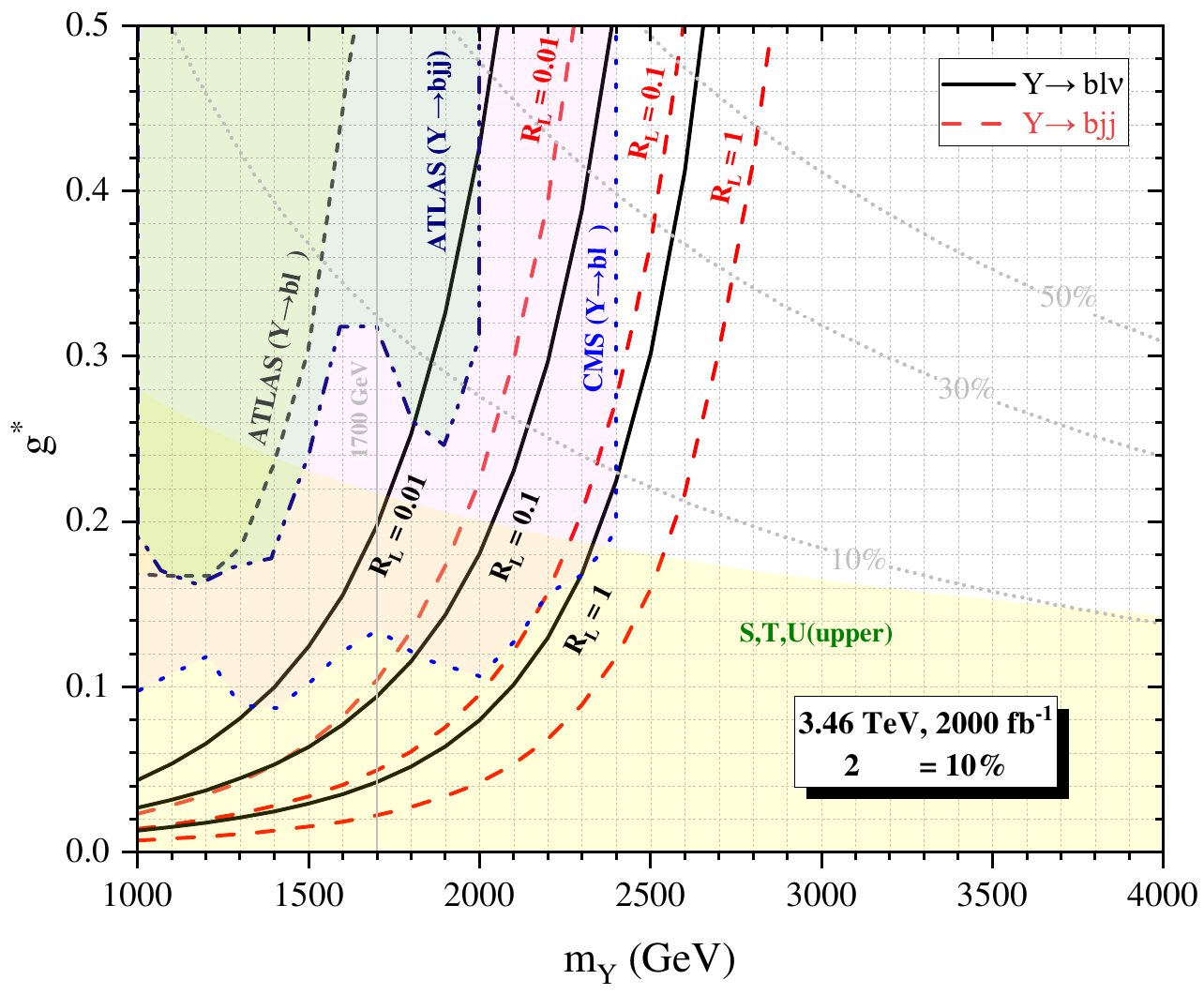}%
    \hspace{0.1\textwidth}%
    \hspace{-1cm}%
    \includegraphics[width=0.45\textwidth]{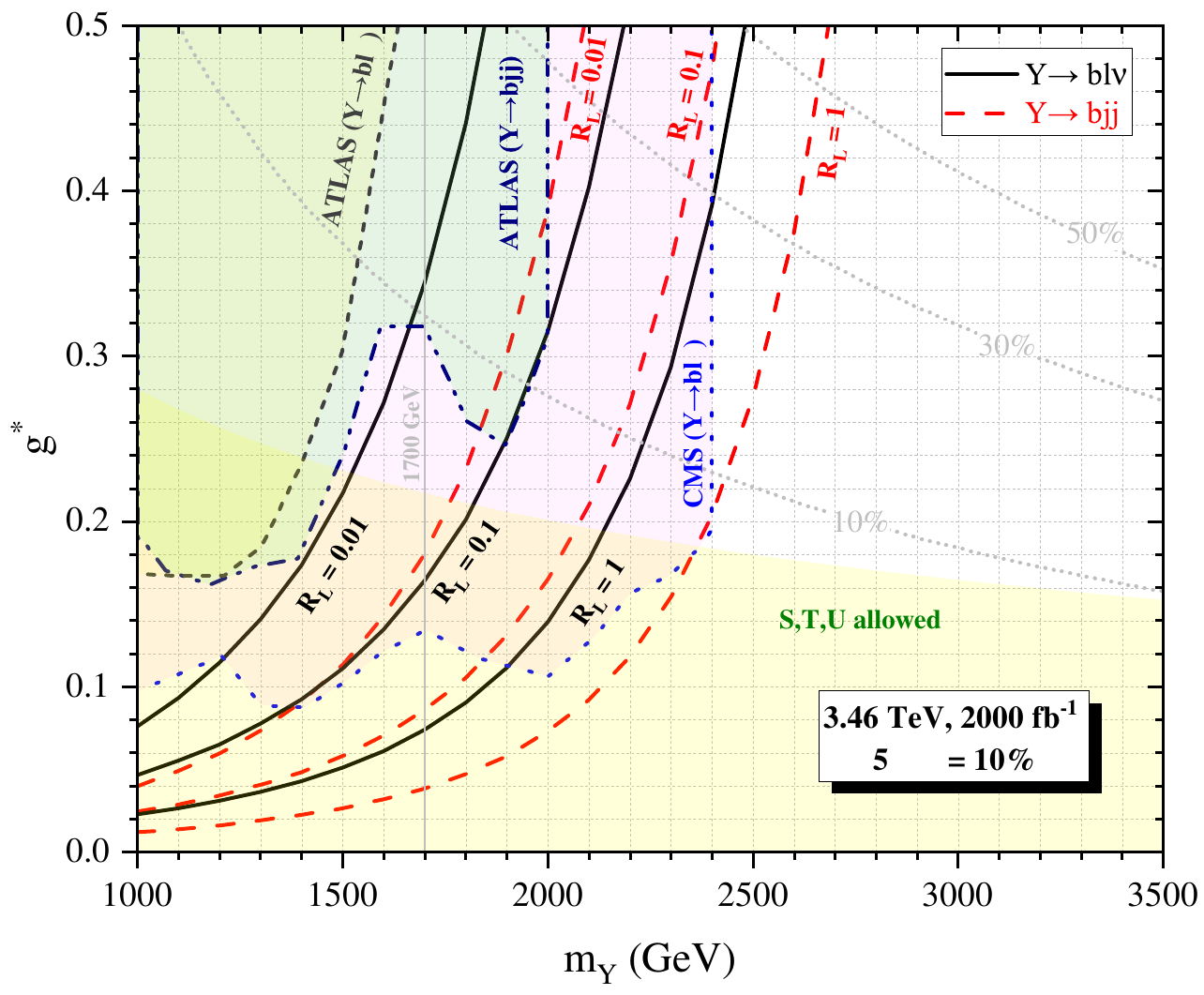}
    
    \vspace{0.2cm}   % 两行之间的垂直间距（可按需调整）
    
    % 第二行
    \includegraphics[width=0.45\textwidth]{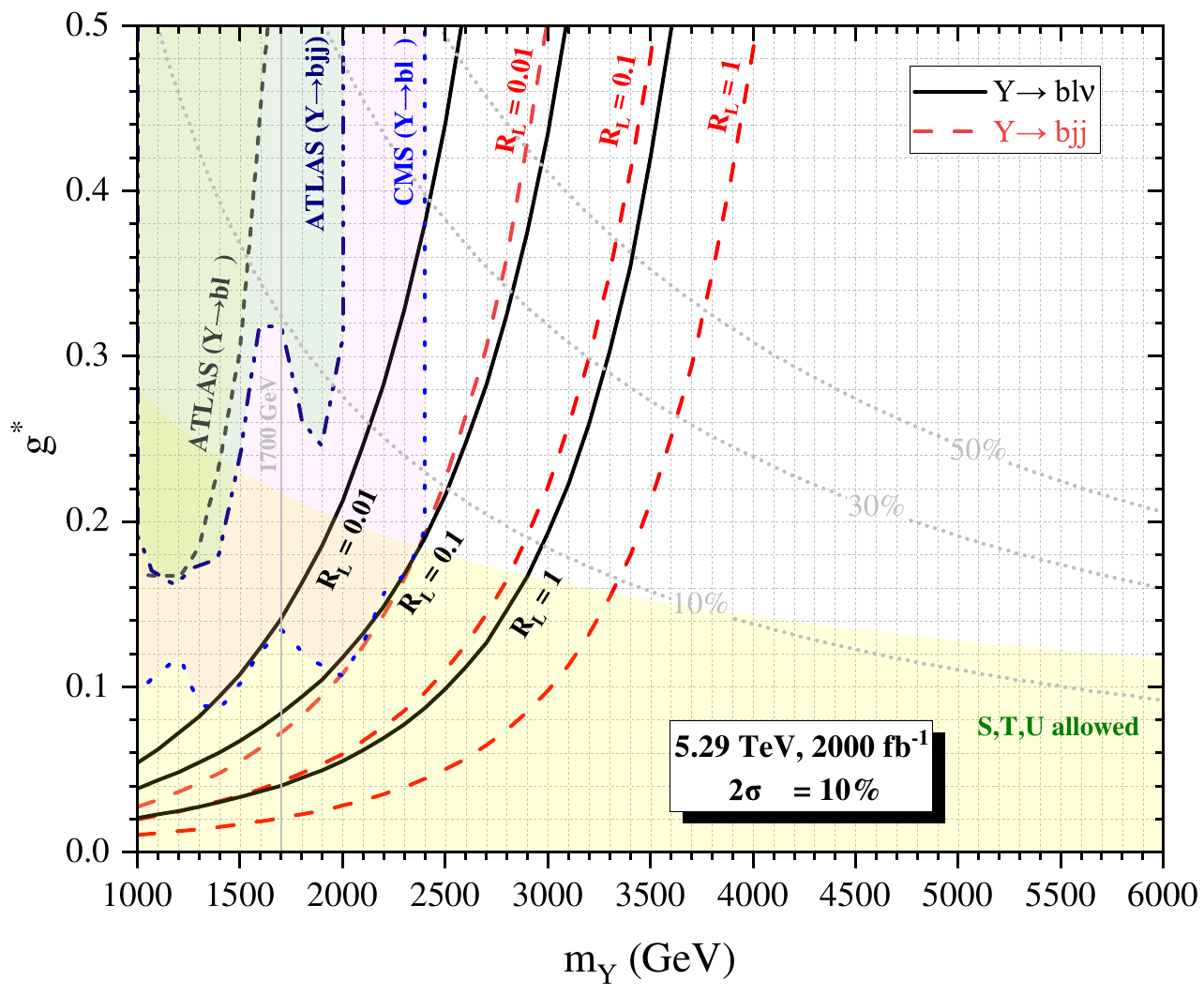}%
    \hspace{0.1\textwidth}%
    \hspace{-1cm}%
    \includegraphics[width=0.45\textwidth]{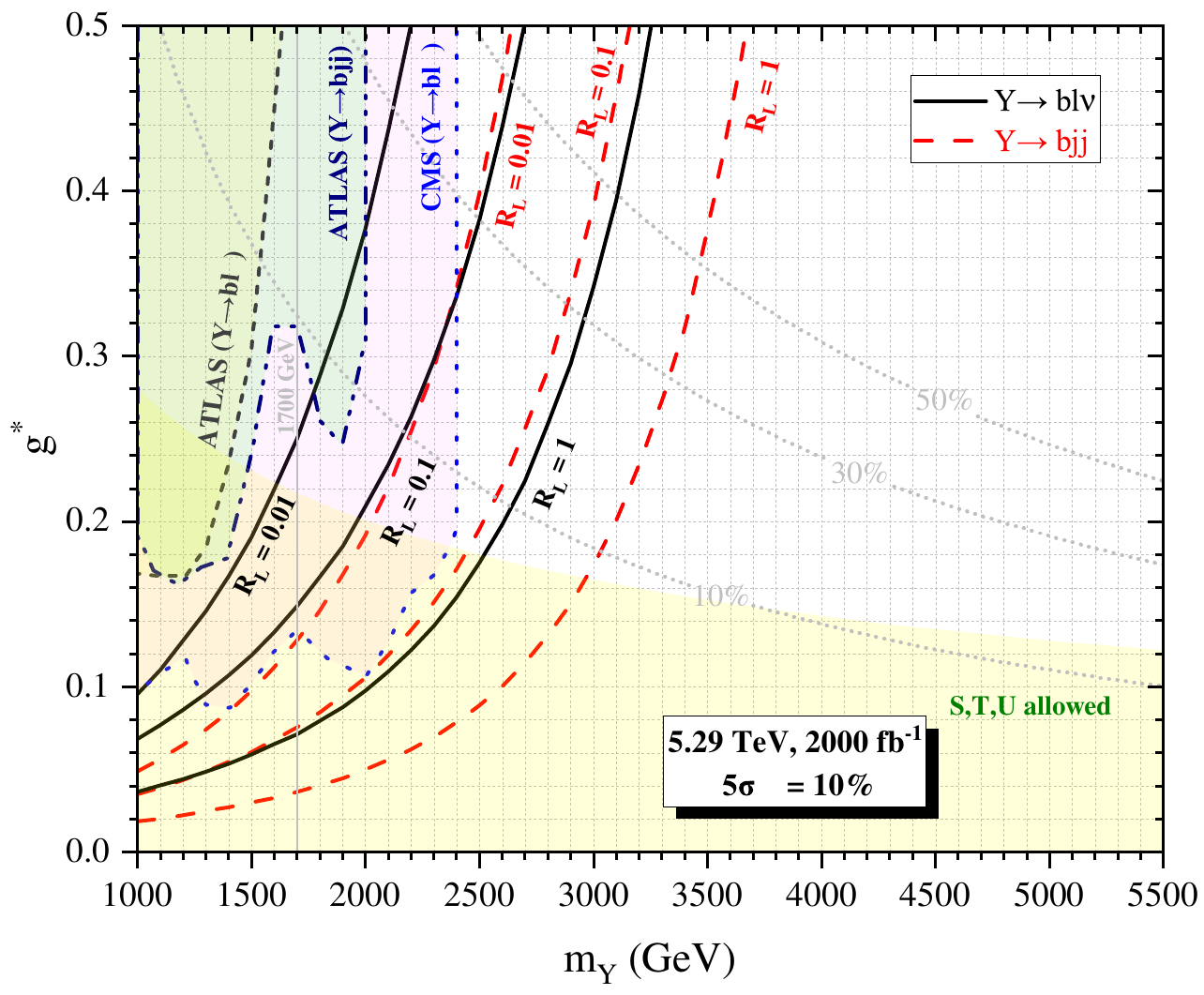}
    
    \vspace{0.2cm}

    % 第三行
    \includegraphics[width=0.45\textwidth]{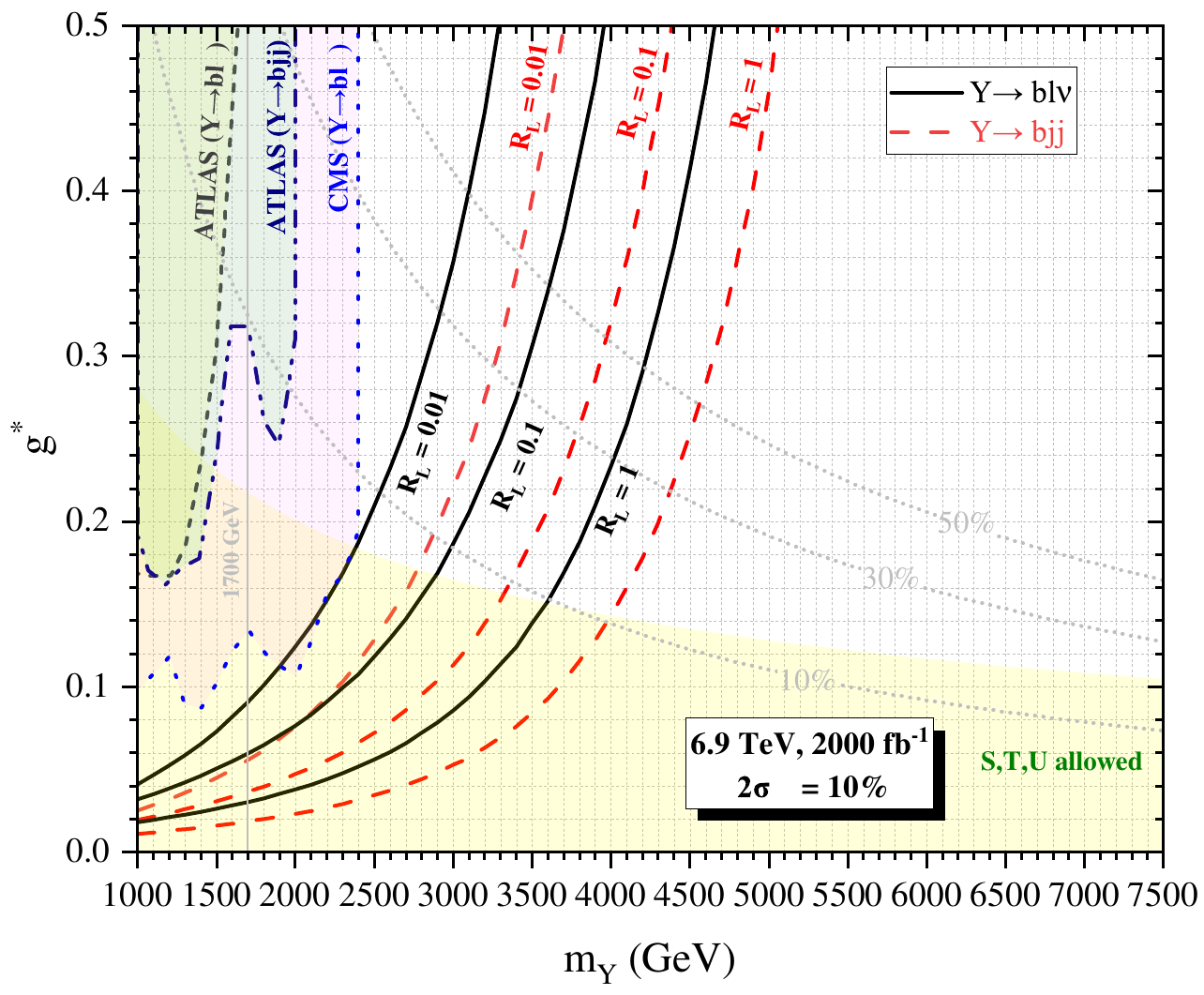}%
    \hspace{0.1\textwidth}%
    \hspace{-1cm}%
    \includegraphics[width=0.45\textwidth]{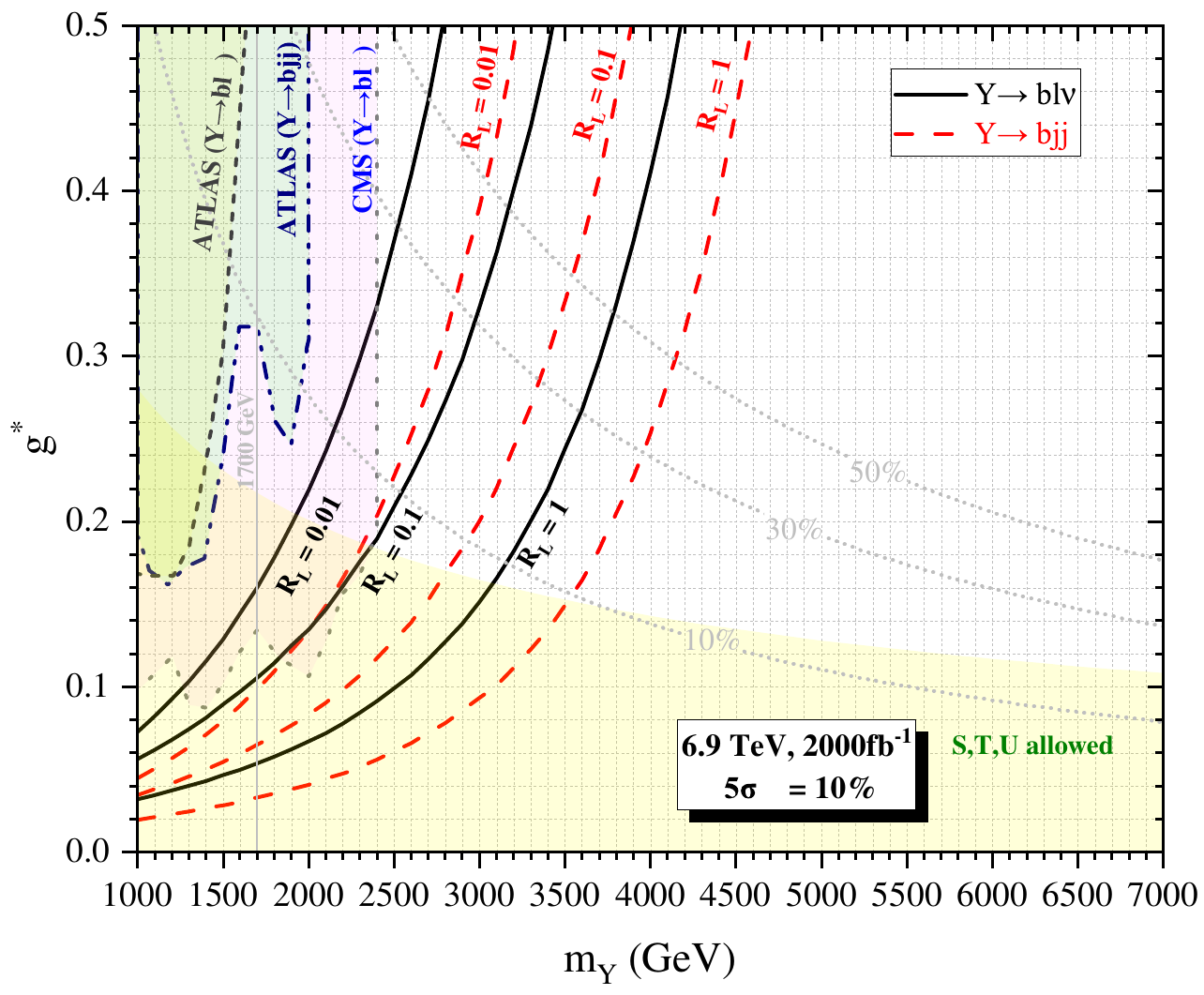}

    \caption{Same as Fig.~\ref{fig6}, but for a systematic uncertainty of $\delta=10\%$.}
    \label{fig7}
\end{figure}

In Figs.~\ref{fig6} and~\ref{fig7}, we show the exclusion and discovery prospects for the $Y$ quark at the FCC-eh with $\sqrt{s}=3.46$, $5.29$, and $6.9~\mathrm{TeV}$ for $R_L=0.01$, $0.1$, and $1$, where the systematic uncertainty of $\delta=0$ and $\delta=10\%$ are displayed, respectively. We also show the existing ATLAS and CMS exclusion limits for the $Y\to b \ell^- \bar{\nu}_{\ell}$~\cite{ATLAS:2018dyh} and $Y\to bjj$~\cite{ATLAS:2024kgp} final states in Figs.~\ref{fig6} and~\ref{fig7}. The vertical gray line denotes the LHC pair-production mass limit, $m_Y>1700~\mathrm{GeV}$~\cite{2024138743}. The yellow shaded region denotes the parameter space allowed by the oblique parameters $S$, $T$, and $U$ using the current fit values in Ref.~\cite{ParticleDataGroup:2024cfk}. Since our simulation is performed in the narrow-width approximation (NWA), where the width-to-mass ratio $\Gamma_Y/m_Y < 10\%$, regions with large $\Gamma_Y/m_Y$ should be interpreted with caution~\cite{PhysRevD.96.075035,PhysRevD.98.015029,Deandrea2021,Moretti:2025ckw}; the contours $\Gamma_Y/m_Y=30\%,50\%$ are shown only as reference lines.

%%%%%%%%%%%%%%%%%%%%%%%%%%%%%%%%%%%%%%%%
%%%%%%%%%%%%%%%%%%%%%%%%%%%%%%%%%%%%%%%%%%%%

%%%%%%%%%%%%%%%%%%%%%%%%%%%%%%%%%%%%%%%%
%%%%%%%%%%%%%%%%%%%%%%%%%%%%%%%%%%%%%%%%%%%%

From Fig.~\ref{fig6}, for Case~1, we can see that the exclusion region in the upper-left panel covers $g^*\in[0.13,0.5]$ and $m_Y\in[1700,2210]~\mathrm{GeV}$ corresponding to $\sqrt{s}=3.46$ TeV and $R_L=0.01$. In the middle-left and lower-left panels, corresponding to $\sqrt{s}=5.29$ and $6.9\mathrm{TeV}$, respectively, the exclusion sensitivity can be improved to $g^*=0.05$ and $0.03$, with mass reaches of $m_Y=3150$ and $4100~\mathrm{GeV}$. For $R_L=1$, the exclusion sensitivity reaches $g^*=0.03$, $0.02$, and $0.01$, with mass reaches of $m_Y=2800$, $4200$, and $5550~\mathrm{GeV}$ for $\sqrt{s}=3.46$, $5.29$, and $6.9\mathrm{TeV}$, respectively. For the same $R_L$, the discovery sensitivity in the upper-right, middle-right, and lower-right panels reaches $g^*=0.04$, $0.02$, and $0.02$, with mass reaches of $m_Y=2650$, $3900$, and $5100~\mathrm{GeV}$.

For Case~2 with $R_L=0.01$, the exclusion sensitivities in the upper-left, middle-left, and lower-left panels are $g^*=0.08$, $0.03$, and $0.02$, with mass reaches of $m_Y=2390$, $3450$, and $4500~\mathrm{GeV}$, respectively. For $R_L=1$, the exclusion sensitivity reaches $g^*=0.02$, $0.01$, and $0.01$, with mass reaches of $m_Y=2990$, $4650$, and $6300~\mathrm{GeV}$. For the same $R_L$, the discovery sensitivity in the upper-right, middle-right, and lower-right panels reaches $g^*=0.03$, $0.01$, and $0.01$, with mass reaches of $m_Y=2800$, $4210$, and $5610~\mathrm{GeV}$. After imposing the APV bound $g^*\lesssim0.13$ for $R_L=1$, the exclusion reach remains $m_Y=2510$, $3700$, and $4700~\mathrm{GeV}$. When $\Gamma_Y/m_Y=10\%$, the corresponding reach becomes $m_Y=2690$, $3750$, and $5100~\mathrm{GeV}$. For $R_L=1$, the APV bound is more restrictive at low masses, while the oblique-parameter constraints from $S$, $T$, and $U$ dominate for $m_Y\gtrsim4900~\mathrm{GeV}$.

Comparing Case~1 and Case~2, we can see that Case~2 provides stronger exclusion and discovery sensitivities for the $Y$ quark than Case~1, and that the sensitivity improves with increasing $\sqrt{s}$ at the FCC-eh.

From Fig.~\ref{fig7}, we can see that the results including a systematic uncertainty of $\delta=10\%$ weaken the sensitivities. Similarly, Case~2 can provide better sensitivity than Case~1, so we only describe Case~2 here. For Case~2 with $R_L=1$, the exclusion region in the upper-left panel covers $g^*\in[0.02,0.5]$ and $m_Y\in[1700,2850]~\mathrm{GeV}$, while the discovery region in the upper-right panel covers $g^*\in[0.04,0.5]$ and $m_Y\in[1700,2680]~\mathrm{GeV}$. In the middle-left panel, the exclusion region covers $g^*\in[0.02,0.5]$ and $m_Y\in[1700,4000]~\mathrm{GeV}$, while the discovery region in the middle-right panel covers $g^*\in[0.04,0.5]$ and $m_Y\in[1700,3690]~\mathrm{GeV}$. In the lower-left panel, the exclusion region covers $g^*\in[0.02,0.5]$ and $m_Y\in[1700,5100]~\mathrm{GeV}$, while the discovery region in the lower-right panel covers $g^*\in[0.03,0.5]$ and $m_Y\in[1700,4600]~\mathrm{GeV}$. Although $R_L=1$ is more strongly constrained by APV, Case~2 at $\sqrt{s}=6.9~\mathrm{TeV}$ still reaches $m_Y=3900~\mathrm{GeV}$ for exclusion in the lower-left panel and $m_Y=3350~\mathrm{GeV}$ for discovery in the lower-right panel. For clarity, we provide further details on the exclusion and discovery reaches of $Y$ quark in Tab.~\ref{table7}.

\newcommand{\CaseOneCell}[1]{\multicolumn{1}{c|}{#1}}

\begin{table}[htbp]
    \centering
    \caption{Summary of the exclusion and discovery reaches of $Y$ quark, where the values outside and inside the brackets correspond to $\delta=0$ (10\%), and the unit of $m_Y$ is GeV.}
    \vspace{0.8cm}

    \resizebox{\textwidth}{!}{%
    \begin{tabular}{l c c c c}
    \toprule[1pt]
    $\sqrt{s}$ (TeV) & $R_L$ & Potential & Case~1 & Case~2 \\
    \midrule[1pt]

    \multirow{6}{*}{3.46}   
    & \multirow{2}{*}{0.01} 
    & Exclusion 
    & \CaseOneCell{$g^* \in$ [0.13 (0.2),0.5] $m_Y \in$ [1700,2210 (2050)]}
    & $g^* \in$ [0.08 (0.1),0.5] $m_Y \in$ [1700,2390 (2290)] \\

    && Discovery  
    & \CaseOneCell{$g^* \in$ [0.2 (0.3),0.5] $m_Y \in$ [1700,2050 (1850)]}
    & $g^* \in$ [0.12 (0.18),0.5] $m_Y \in$ [1700,2200 (2090)] \\ 
    \cline{2-5}

    & \multirow{2}{*}{0.1}
    & Exclusion 
    & \CaseOneCell{$g^* \in$ [0.06 (0.09),0.5] $m_Y \in$ [1700,2525 (2390)]}
    & $g^* \in$ [0.04 (0.05),0.5] $m_Y \in$ [1700,2700 (2600)] \\

    && Discovery  
    & \CaseOneCell{$g^* \in$ [0.1 (0.16),0.5] $m_Y \in$ [1700,2390 (2190)]}
    & $g^* \in$ [0.06 (0.09),0.5] $m_Y \in$ [1700,2550 (2400)] \\  
    \cline{2-5}

    & \multirow{2}{*}{1}
    & Exclusion 
    & \CaseOneCell{$g^* \in$ [0.03 (0.04),0.5] $m_Y \in$ [1700,2800 (2650)]}
    & $g^* \in$ [0.02 (0.02),0.5] $m_Y \in$ [1700,2990 (2850)] \\

    && Discovery  
    & \CaseOneCell{$g^* \in$ [0.04 (0.07),0.5] $m_Y \in$ [1700,2650 (2475)]}
    & $g^* \in$ [0.03 (0.04),0.5] $m_Y \in$ [1700,2800 (2680)] \\

    \midrule[1pt]

    \multirow{6}{*}{5.29}   
    & \multirow{2}{*}{0.01} 
    & Exclusion 
    & \CaseOneCell{$g^* \in$ [0.05 (0.14),0.5] $m_Y \in$ [1700,3150 (2590)]}
    & $g^* \in$ [0.03 (0.07),0.5] $m_Y \in$ [1700,3450 (3000)] \\

    && Discovery  
    & \CaseOneCell{$g^* \in$ [0.09 (0.25),0.5] $m_Y \in$ [1700,2900 (2200)]}
    & $g^* \in$ [0.05 (0.13),0.5] $m_Y \in$ [1700,3200 (2650)] \\ 
    \cline{2-5}

    & \multirow{2}{*}{0.1}
    & Exclusion 
    & \CaseOneCell{$g^* \in$ [0.03 (0.08),0.5] $m_Y \in$ [1700,3700 (3100)]}
    & $g^* \in$ [0.02 (0.04),0.5] $m_Y \in$ [1700,4000 (3500)] \\

    && Discovery  
    & \CaseOneCell{$g^* \in$ [0.05 (0.15),0.5] $m_Y \in$ [1700,3400 (2700)]}
    & $g^* \in$ [0.03 (0.08),0.5] $m_Y \in$ [1700,3700 (3150)] \\  
    \cline{2-5}

    & \multirow{2}{*}{1}
    & Exclusion 
    & \CaseOneCell{$g^* \in$ [0.02 (0.04),0.5] $m_Y \in$ [1700,4200 (3600)]}
    & $g^* \in$ [0.01 (0.02),0.5] $m_Y \in$ [1700,4650 (4000)] \\

    && Discovery  
    & \CaseOneCell{$g^* \in$ [0.02 (0.07),0.5] $m_Y \in$ [1700,3900 (3250)]}
    & $g^* \in$ [0.01 (0.04),0.5] $m_Y \in$ [1700,4210 (3690)] \\

    \midrule[1pt]

    \multirow{6}{*}{6.9}   
    & \multirow{2}{*}{0.01} 
    & Exclusion 
    & \CaseOneCell{$g^* \in$ [0.03 (0.09),0.5] $m_Y \in$ [1700,4100 (3300)]}
    & $g^* \in$ [0.02 (0.06),0.5] $m_Y \in$ [1700,4500 (3700)] \\

    && Discovery  
    & \CaseOneCell{$g^* \in$ [0.05 (0.16),0.5] $m_Y \in$ [1700,3700 (2800)]}
    & $g^* \in$ [0.03 (0.1),0.5] $m_Y \in$ [1700,4100 (3200)] \\ 
    \cline{2-5}

    & \multirow{2}{*}{0.1}
    & Exclusion 
    & \CaseOneCell{$g^* \in$ [0.02 (0.06),0.5] $m_Y \in$ [1700,4800 (3950)]}
    & $g^* \in$ [0.01 (0.04),0.5] $m_Y \in$ [1700,5300 (4400)] \\

    && Discovery  
    & \CaseOneCell{$g^* \in$ [0.04 (0.1),0.5] $m_Y \in$ [1700,4400 (3400)]}
    & $g^* \in$ [0.02 (0.07),0.5] $m_Y \in$ [1700,4800 (3900)] \\  
    \cline{2-5}

    & \multirow{2}{*}{1}
    & Exclusion 
    & \CaseOneCell{$g^* \in$ [0.01 (0.03),0.5] $m_Y \in$ [1700,5550 (4650)]}
    & $g^* \in$ [0.01 (0.02),0.5] $m_Y \in$ [1700,6300 (5100)] \\

    && Discovery  
    & \CaseOneCell{$g^* \in$ [0.02 (0.05),0.5] $m_Y \in$ [1700,5100 (4190)]}
    & $g^* \in$ [0.01 (0.03),0.5] $m_Y \in$ [1700,5610 (4600)] \\

    \bottomrule[1pt]
    \end{tabular}%
    }

    \label{table7}
\end{table}

\section{Conclusions}
\label{section5}

In this work, we studied the single production of a vector-like $Y$ quark via $e^-p\to Y (\to bW)\nu_e$ at the FCC-eh. The analysis was carried out for $P_e=-80\%$ at $\sqrt{s}=3.46$, $5.29$, and $6.9\mathrm{TeV}$, considering both the leptonic final state $Y\to bW(\to \ell^-\bar{\nu}_{\ell})$ and the hadronic final state $Y\to bW(\to jj)$. After a detector-level simulation, cut-based analyses were performed for the two channels, and the corresponding $2\sigma$ exclusion limits and $5\sigma$ discovery reaches were obtained in the $g^*$--$m_Y$ plane for different choices of $R_L$. The obtained sensitivities were also compared with the current LHC constraints. Given the $t$-channel structure of the production process, we found that the sensitivity is significantly improved at higher collision energies. In particular, Case~2 (hadronic channel), in which the hadronically decaying $W$-boson is reconstructed as a $W$-jet, yields better exclusion and discovery sensitivities than Case~1 (leptonic channel). These results indicate that the FCC-eh can provide a complementary probe of vector-like $Y$ quarks beyond the LHC.

Ref.~\cite{Shang:2024wwy} studied single production of a doublet $Y$ quark in the $Y\to bW (\to \ell^- \bar{\nu}_{\ell})$ channel at the $14~\mathrm{TeV}$ HL-LHC with $3000~\mathrm{fb}^{-1}$ and a systematic uncertainty of $\delta=10\%$, obtaining exclusion and discovery regions of $\kappa_Y\in[0.05,0.5]$, $m_Y\in[1500,3970]~\mathrm{GeV}$, and $\kappa_Y\in[0.09,0.5]$, $m_Y\in[1500,3360]~\mathrm{GeV}$, respectively. Ref.~\cite{CETINKAYA2021115580} considered the $Y\to bW (\to jj)$ channel at the same collider configuration and obtained exclusion and discovery reaches of $m_Y=2350~\mathrm{GeV}$ and $1900~\mathrm{GeV}$ for $\kappa_Y=0.5$, respectively. Since $g^*=\kappa_Y$, our results indicate that the FCC-eh can extend the sensitivity to smaller couplings.

\begin{acknowledgments}
S.M. is supported in part through the NExT Institute and STFC Consolidated Grant ST/X000583/1. 
W.Z. thanks Jin Min Yang for helpful discussions. This work is supported by the Henan Provincial International Science, Technology Cooperation Incubation Project 262102520027 and the High Performance Computing Platform of Henan Normal University.
\end{acknowledgments}

\bibliography{myref}  %不能放在导言区
\end{document}